\documentclass[onecolumn,prd,amssymb,amsmath,preprintnumbers,aps,nofootinbib,superscriptaddress,11pt]{revtex4-1}
\usepackage[a4paper, portrait, margin=0.7in]{geometry}
\usepackage{lineno}
\usepackage{bm,color,xcolor}
\usepackage{slashed} 
\usepackage{graphicx}
\usepackage{setspace}
\usepackage{soul} 
\usepackage{multirow}
\usepackage{comment}
\usepackage{epstopdf} 
\usepackage{mathtools}
\usepackage{draftfigure}
\usepackage{appendix}
\usepackage[colorlinks=true
,urlcolor=blue
,anchorcolor=blue
,citecolor=blue 
,filecolor=blue
,linkcolor=blue
,menucolor=blue
,linktocpage=true
,pdfproducer=medialab
,pdfa=true
]{hyperref}

\newcommand{\beq}{\begin{equation}}
\newcommand{\eeq}{\end{equation}}
\newcommand{\bea}{\begin{eqnarray}}
\newcommand{\eea}{\end{eqnarray}}
\newcommand{\ba}{\begin{array}}
\newcommand{\ea}{\end{array}}

\def\m1{M_1}
\def\m2{M_2}
\def\m3{M_3}

\def\ch10{\tilde \chi^0_1}

\def\gev{\,{\rm GeV}}

\def\to{\rightarrow}

\newcommand{\lsim}{\mathrel{\mathop{\kern 0pt \rlap
  {\raise.2ex\hbox{$<$}}}
  \lower.9ex\hbox{\kern-.190em $\sim$}}}
\newcommand{\gsim}{\mathrel{\mathop{\kern 0pt \rlap
  {\raise.2ex\hbox{$>$}}}
  \lower.9ex\hbox{\kern-.190em $\sim$}}}
\begin{document}
\title{\boldmath \bf \Large 
Heavy Axion Opportunities at the DUNE Near Detector
}

\author{Kevin J. Kelly}
\email{kkelly12@fnal.gov}
\thanks{\scriptsize \!\! \href{https://orcid.org/0000-0002-4892-2093}{0000-0002-4892-2093}}
\affiliation{Theoretical Physics Department, Fermi National Accelerator Laboratory, P. O. Box 500, Batavia, IL 60510, USA}
\author{Soubhik Kumar}
\email{soubhik@berkeley.edu}
\thanks{\scriptsize \!\! \href{https://orcid.org/0000-0001-6924-3375}{0000-0001-6924-3375}}
\affiliation{Berkeley Center for Theoretical Physics, Department of Physics,
University of California, Berkeley, CA 94720, USA}
\affiliation{Theoretical Physics Group, Lawrence Berkeley National Laboratory, Berkeley, CA 94720, USA}
\affiliation{Maryland Center for Fundamental Physics, Department of Physics, University of Maryland, College Park, MD 20742, USA}
\author{Zhen Liu}
\email{zliuphys@umn.edu}
\thanks{\scriptsize \!\! \href{https://orcid.org/0000-0002-3143-1976}{0000-0002-3143-1976}}
%\thanks{\scriptsize \!\! \href{https://orcid.org/0000-0001-6924-3375}{0000-0001-6924-3375}}
\affiliation{Maryland Center for Fundamental Physics, Department of Physics, University of Maryland, College Park, MD 20742, USA}
\affiliation{School of Physics and Astronomy, University of Minnesota, Minneapolis, MN 55455, USA}
\preprint{FERMILAB-PUB-20-581-T}
\date{\normalsize  \today}
\setstretch{1.0}

\begin{abstract}
While the QCD axion is often considered to be necessarily light ($\lesssim$ eV), recent work has opened a viable and interesting parameter space for heavy axions, which solve both the Strong CP and the axion Quality Problems. These well-motivated heavy axions, as well as the generic axion-like-particles, call for explorations in the GeV mass realm at collider and beam dump environments.
The primary upcoming neutrino experiment, Deep Underground Neutrino Experiment (DUNE), is simultaneously also a powerful beam dump experiment, enabled by its multipurpose Near Detector (ND) complex. In this study, we show with detailed analyses that the DUNE ND has a unique sensitivity to heavy axions for masses between $20$~MeV and $2$~GeV, complementary to other future experiments.
\end{abstract}

\preprint{
}

{
\let\clearpage\relax
\maketitle
}
%\linenumbers
\tableofcontents
\section{Introduction}
\label{sec:intro}

The QCD axion provides an elegant dynamical solution to the Strong CP Problem of the Standard Model (SM)~\cite{Peccei:1977hh,Peccei:1977ur,Weinberg:1977ma,Wilczek:1977pj}.
In its minimal realization, the QCD axion signatures are dominated by its couplings to the CP-odd field strength operators of SM gauge fields and the SM matter fields. The well-known relation (see e.g.~\cite{diCortona:2015ldu}) for the QCD axion, relating its mass $m_a$ to its decay constant $f_a$ -- $m_a=5.7\left(\frac{10^9\ \rm{GeV}}{f_a}\right)\rm{meV}$ -- along with a variety of experimental/observational constraints implying $f_a>10^{9}$~GeV~\cite{Zyla:2020zbs}, have driven most QCD axion searches to focus on light, sub-eV masses. However, a series of recent model building efforts~\cite{Hook:2014cda,Fukuda:2015ana,Dimopoulos:2016lvn,Gherghetta:2016fhp,Agrawal:2017evu,Agrawal:2017ksf,Lillard:2018fdt,Gaillard:2018xgk,Hook:2019qoh,Csaki:2019vte,Gherghetta:2020keg} including earlier work~\cite{Dimopoulos:1979pp,Holdom:1982ex,Dine:1981rt,Flynn:1987rs,Rubakov:1997vp,Berezhiani:2000gh} motivate heavier variants of the QCD axion and, within a class of such models relying on a $Z_2-$symmetric mirror SM sector, a testable parameter space has been identified where the axion mass can be around or even larger than the GeV scale~\cite{Hook:2019qoh}. In such scenarios, the heavy axion can be obtained by introducing a new strongly-coupled mirror $SU(3)$ sector that also generates a larger axion potential aligned with the QCD-generated potential~\cite{Rubakov:1997vp,Berezhiani:2000gh,Hook:2014cda,Fukuda:2015ana,Dimopoulos:2016lvn,Hook:2019qoh}. As a result the axion continues to solve the Strong CP problem, while being heavy enough (for a given $f_a$) to be more robust against unwanted UV contributions which would otherwise have given rise to the so-called Quality Problem~\cite{Kamionkowski:1992mf,Barr:1992qq,GHIGNA1992278,Holman:1992us}. Therefore, this predicts a new, less-explored heavy axion solving \textit{both} the Strong CP and the Quality Problem that deems further exploration. Moreover, a pseudoscalar field is a generic constituent of many beyond-the-Standard-Model (BSM) scenarios~\cite{Beacham:2019nyx}, as well as String Theoretic constructions~\cite{Svrcek:2006yi,Arvanitaki:2009fg}. Hence, the search for GeV-scale pseudoscalar fields, parametrized under a generic effective field theory, dubbed as ``Axion-like Particles'' (ALP), is a vital component of the BSM program.

Existing theoretical studies show that such heavy axions can be probed at beam dump and fixed target experiments%~\cite{Bergsma:1985qz,Bjorken:1988as,Blumlein:1991xh,Adler:2004hp,Artamonov:2009sz,
~\cite{Dobrich:2015jyk,Dobrich:2017gcm,Dolan:2017osp,Dobrich:2019dxc,Harland-Lang:2019zur,Dent:2019ueq,AristizabalSierra:2020rom} along with electron-positron colliders and hadron colliders~\cite{Jaeckel:2012yz,Mimasu:2014nea,Jaeckel:2015jla,Bauer:2017nlg,Bauer:2017ris,Mariotti:2017vtv,Brivio:2017ije,Hook:2019qoh,Ebadi:2019gij,Gavela:2019cmq}. Searches at these different facilities complement each other and their sensitivities have become better understood thanks to many recent developments in the understanding of the properties of the heavy axion~\cite{Bauer:2017nlg,Bauer:2017ris,Bauer:2019gfk,Aloni:2018vki} and their production via different mechanisms~\cite{Izaguirre:2016dfi,Knapen:2017ebd,CidVidal:2018blh,Aloni:2019ruo,Altmannshofer:2019yji,Gori:2020xvq,Brdar:2020dpr}. Various future experiments would also be able to probe interesting parts of the axion parameter space~\cite{Chou:2016lxi,Gligorov:2017nwh,Feng:2018noy,Bakhti:2020vfq}. While many of the existing searches focus on axion-photon or axion-electroweak couplings, a solution to the Strong CP problem motivates an equally or more dominant axion-gluon coupling and in that regard, proton beam dump experiments and hadron colliders can play a crucial role.

The next-generation Deep Underground Neutrino Experiment (DUNE)~\cite{Abi:2020evt}, via its intense, high energy Long Baseline Neutrino Facility (LBNF) proton beam and high-precision near and far detectors, will facilitate the most accurate measurement of various neutrino properties. At the same time, its Near Detector (ND) site~\cite{Abi:2020evt}, composed of a complex detector facility with a full-fledged particle detection and identification system, can be creatively viewed as a beam dump facility for new physics searches for long-lived particle signatures~\cite{Ballett:2019bgd,Berryman:2019dme,Coloma:2020lgy,Abi:2020kei}. As we demonstrate in this work, the heavy axions, well-motivated by the Strong CP and Quality Problems, and generic BSM considerations, would provide a compelling target for DUNE ND.  Our proposed search at DUNE ND can cover unique parts of the axion parameter space.

\begin{figure}
\begin{center}
\includegraphics[width=0.85\linewidth]{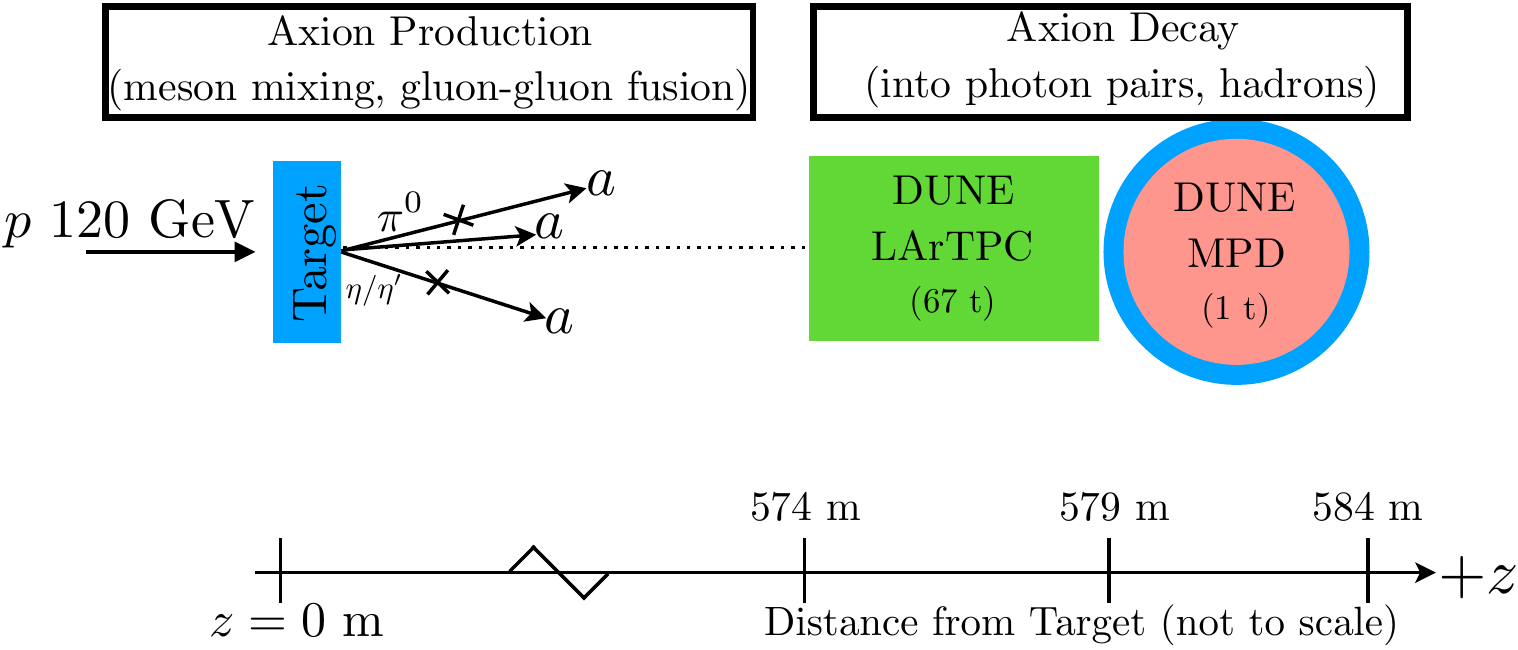}
\caption{Simplified schematic representation of the LBNF beam, DUNE target, and DUNE Near Detector complex for our study. Protons with $120$ GeV energy strike a target that produces copious amounts of SM mesons and, potentially, axions $a$. Some fraction of these particles travel in the direction of the Near Detector complex (right) where the liquid argon time projection chamber (green) and multi-purpose detector (red indicating the gaseous argon time-projection chamber and blue indicating the electromagnetic calorimeter/magnet) are situated. The axions can decay in those detectors providing striking signals. The crosses on the $\pi^0$ and $\eta/\eta^\prime$ lines indicate that these particles can directly mix with the axions $a$. Figure adapted and modified from Ref.~\cite{Berryman:2019dme}.
\label{fig:NDDrawing}}
\end{center}
\end{figure}

This work is organized as follows. In Sec.~\ref{sec.heavyax}, we briefly review the mechanism solving both the Strong CP and Quality Problems and how that motivates a heavy axion. After discussing some of its properties and the EFT parametrization, in Sec.~\ref{sec.axionsim}, we detail our simulation procedure for axion production in the proton beam dump environment with the DUNE ND. That allows us to investigate in Sec.~\ref{sec:ExpSignatures} both the signatures of heavy axion decay in DUNE ND and possible background contributions. All of these considerations then enable us to derive the DUNE ND sensitivity to heavy axions in Sec.~\ref{sec.NDsens}. When projecting sensitivities, we consider two benchmark models, both orthogonal to the ``photon-dominant'' axion(-like-particles) frequently considered in the literature. We assume that either (a) the heavy axion coupling is ``gluon-dominated'', i.e., the coupling between the axion and the QCD field strength tensor is the largest among its other couplings, or (b) the heavy axion coupling is ``co-dominant'' and the axion couples equally to the different SM field strength tensors of $SU(3)$, $SU(2)_W$ and $U(1)_Y$. We then conclude in Sec.~\ref{sec.conclusion}.

\section{Properties of a Heavy Axion}\label{sec.heavyax}
This section details the properties of a heavy axion considered in this work. First, in section~\ref{subsec:StrongCP}, we introduce such a heavy axion that solves the Strong CP Problem. After discussing the axion Quality Problem, in Section~\ref{subsec:QualityProb}, we demonstrate how such a heavy axion can solve both these problems simultaneously and thereby be theoretically well-motivated. After reviewing the theoretical constraints on such a heavy axion, in Section~\ref{subsec:EFT}, we describe the axion properties under benchmark choices of the axion effective field theory (EFT) to study experimental prospects in the later sections.

Before proceeding, we wish to clarify some definitions and conventions that will be used throughout this work. The (dimensionful) coupling $f_a$ is the axion decay constant that enters as a part of axion coupling to the CP-odd QCD field strength tensor, responsible for solving the Strong CP problem. We will also occasionally use the quantity $f_G = 4\pi^2 f_a$, which is adopted in some phenomenological studies of heavy axions in the literature. Finally, since production rates and decay widths of $a$ are typically inversely-proportional to $f_a$ and $f_G$, we will often use the coupling $g_{agg} = f_G^{-1}$ so that these rates/widths are proportional to positive powers of $g_{agg}$, which also allows for more transparent comparisons between our projections and those in the literature.

\subsection{The Strong CP and the Quality Problem}\label{subsec:StrongCP}
As defined, the QCD axion has a gluonic coupling (for a recent review see \cite{DiLuzio:2020wdo})
\begin{eqnarray}
\mathcal{L} \supset \frac{\alpha_3}{8\pi}\left(\bar{\theta}+\frac{a(x)}{f_a}\right)G_{\mu\nu}^a \tilde{G}^{a,\mu\nu},
\end{eqnarray}
through which it solves the Strong CP Problem dynamically. This can be seen explicitly by considering the QCD-generated
axion potential \cite{Weinberg:1977ma,DiVecchia:1980yfw,diCortona:2015ldu}
\begin{align}
\mathcal{V}\approx -m_\pi^2 f_\pi^2\sqrt {1-\frac {4m_u m_d} {(m_u+m_d)^2}\sin^2\left(\frac {a(x)} {2f_a}+\frac {\bar\theta} 2\right)}.
\label{eq:QCDV}
\end{align}
As the axion dynamically relaxes to its minima at $\langle a\rangle = -\bar{\theta}f_a$, it makes the \textit{effective} $\bar{\theta}$ parameter vanish -- solving the Strong CP Problem and explaining the smallness of the (as-yet unobserved) neutron electric dipole moment~\cite{Abel:2020gbr}. The above QCD-axion potential gives rise to the well-known relation~\cite{diCortona:2015ldu},
\begin{align}\label{eq.mafaqcd}
m_a=5.7\left(\frac{10^9\ \rm{GeV}}{f_a}\right)\rm{meV}.
\end{align}
However, a variety of terrestrial, astrophysical and cosmological constraints (see e.g. \cite{Vysotsky:1978dc,Raffelt:2006cw,Cadamuro:2011fd,Millea:2015qra,Zyla:2020zbs}) require $f_a > 10^{9}$ GeV for the QCD axion and therefore, the relation in Eq.~\eqref{eq.mafaqcd} precludes observing an otherwise phenomenologically interesting, accelerator-observable parameter space where $m_a\sim 10$ MeV - $100$ GeV. Hence, it is interesting to ask whether there exist models solving the Strong CP Problem which can occupy this mass regime.

At the same time, from an ultraviolet (UV) perspective, various axion models often suffer from the so-called ``Quality Problem''~\cite{Kamionkowski:1992mf,Barr:1992qq,GHIGNA1992278,Holman:1992us}. To see this, we recall that an axion can be realized as the Goldstone boson of a spontaneously broken Peccei-Quinn (PQ) $U(1)_{\rm PQ}$ symmetry. In the far UV, quantum gravitational effects are expected to break all global symmetries~\cite{Kallosh:1995hi,Banks:2010zn}. Therefore,  at low energies, the $U(1)_{\rm PQ}$ can,  at best,  survive as some accidental symmetry.
%\footnote{Unless we promote the PQ symmetry to a local gauge symmetry in the UV.} 
To illustrate the severity of the Quality Problem, we consider a Planck-suppressed $U(1)_{\rm PQ}$-breaking operator,
\begin{align}\label{eq.quality}
\frac{\Phi^N}{M_{\rm pl}^{N-4}} \sim \frac{f_a^N}{M_{\rm pl}^{N-4}} e^{iNa/f_a}.
\end{align}
In the above, the axion $a$ arises as the Goldstone mode of the (composite) field $\Phi\sim f_a e^{ia/f_a}$ at low energies, and the resulting minima-structure of the axion potential need not align with the QCD-generated potential. Therefore, unless the UV contribution in Eq.~\eqref{eq.quality} is small compared to the QCD-generated potential in Eq.~\eqref{eq:QCDV}, the axion will relax to a minima dictated by Eq.~\eqref{eq.quality} where generally $\langle a\rangle\neq -\bar{\theta}f_a$ and the axion solution to the Strong CP Problem will be spoiled. More concretely, given the constraint $f_a>10^{9}$ GeV for the minimal QCD axion, we see that unless we forbid all operators of the type in Eq.~\eqref{eq.quality} up to $N=9$, the axion-solution to the Strong CP Problem no longer works. The question of why the UV theory should respect $U(1)_{\rm PQ}$ to such high quality is the Quality Problem.

\subsection{Addressing the Quality Problem}\label{subsec:QualityProb}
\begin{figure}
\begin{center}
\includegraphics[width=0.8\linewidth]{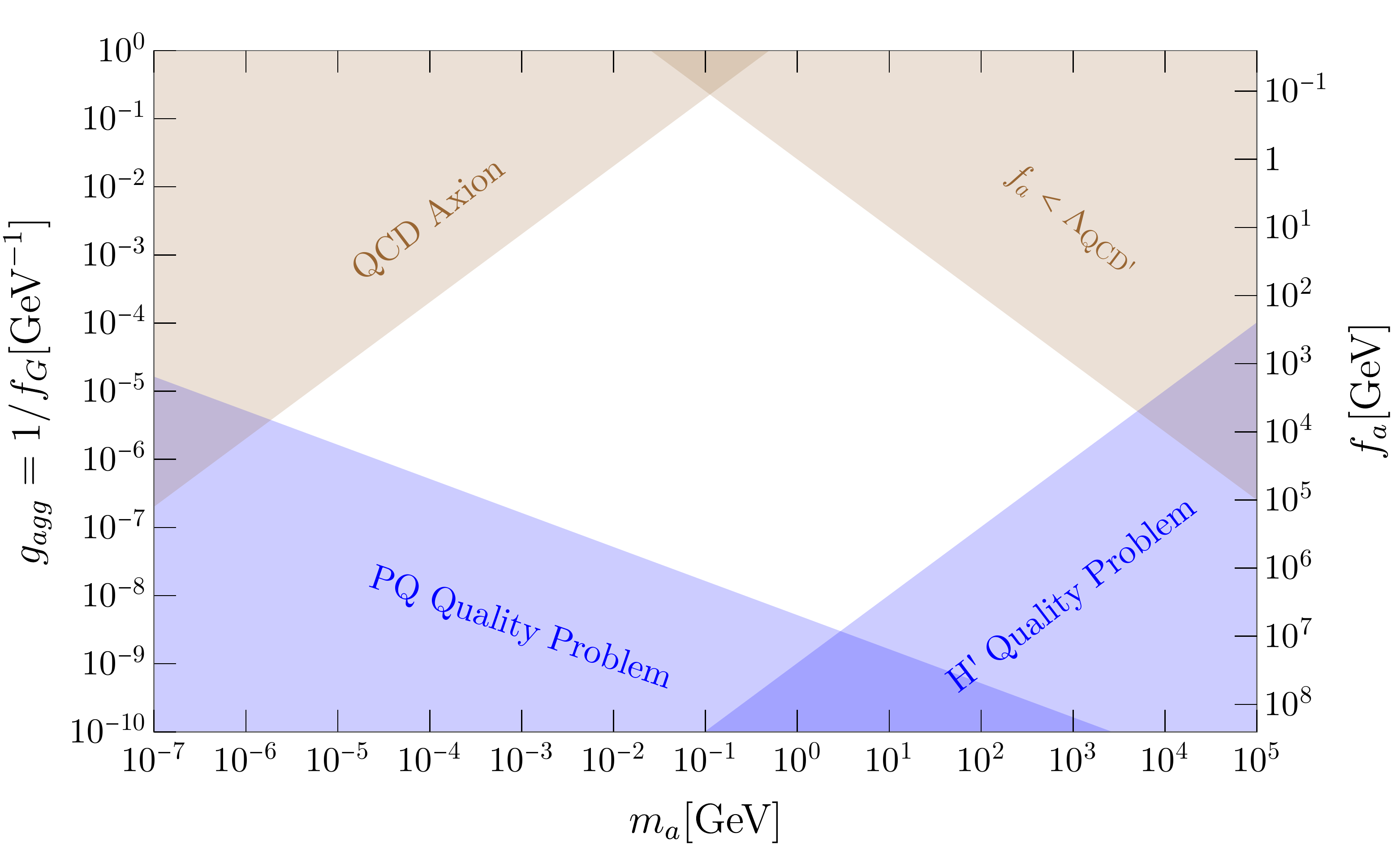}
\caption{Theoretical constraints on the axion parameter space for the class of models considered in this work that solve both the Strong CP and the Quality Problems, adapted from Ref.~\cite{Hook:2019qoh}.
The white region is the theoretically allowed/motivated region. See the text for explanations of different labels. The parameters $f_G$ and $f_a$ are related by $f_G=4\pi^2 f_a$.
}
\label{fig:theoryspace}
\end{center}
\end{figure}

In Ref.~\cite{Hook:2019qoh} a model addressing the Quality Problem was constructed, generalizing on previous work \cite{Rubakov:1997vp,Berezhiani:2000gh,Hook:2014cda,Fukuda:2015ana,Dimopoulos:2016lvn}, in which the Strong CP Problem is solved through the presence of a $Z_2$-symmetric mirror sector (containing the primed fields)\footnote{For other approaches for addressing the Quality Problem, including some recent discussions,  see e.g. ~\cite{Kim:1984pt,Randall:1992ut,Choi:2003wr,Fukuda:2017ylt,Lillard:2018fdt,Gavela:2018paw,Cox:2019rro,Alvey:2020nyh}}. The axion coupling is then given by,
\begin{eqnarray}
\frac{\alpha_3}{8\pi}\left(\bar{\theta}+\frac{a(x)}{f_a}\right)\left(G_{\mu\nu}^a \tilde{G}^{a,\mu\nu}+G_{\mu\nu}^{\prime a} \tilde{G}^{\prime a,\mu\nu}\right).
\end{eqnarray}
This $Z_2$ symmetry is softly broken by the only relevant operator in the SM and the mirror sector, the Higgs masses. Consequently, the mirror Higgs VEV $\langle H'\rangle$ can ``naturally'' be much larger than $\langle H\rangle$. In this case, the mirror quarks decouple at higher energies without impacting the RG running of the mirror QCD at lower energies. This results in the mirror confinement scale $\Lambda_{\rm QCD'}$ being much larger than $\Lambda_{\rm QCD}$. Therefore, the QCD-axion potential receives a parametrically larger contribution from the mirror sector with \cite{Hook:2019qoh},
\begin{align}
\mathcal{V}\supset \Lambda_{\rm QCD'}^{4} \left(\frac{a}{f_a}+\bar{\theta}\right)^2+\cdots.
\end{align}
At the same time, thanks to the $Z_2$ symmetry, this enhanced contribution still aligns with SM-QCD generated potential in Eq.~\eqref{eq:QCDV}. Thus the same axion solves the Strong CP Problem in both the sectors, and, since the axion potential is parametrically enhanced due to the presence of the mirror sector, it is less susceptible to the Quality Problem.  Since the RG running of $\bar{\theta}$ happens at seven loops and the threshold effects happen at four loops~\cite{Ellis:1978hq},  the two $\bar{\theta}$ angles remain approximately equal in the IR despite the spontaneous $Z_2$-breaking. We now briefly summarize the theoretical constraints on our model in Fig.~\ref{fig:theoryspace}, while referring the reader to Ref.~\cite{Hook:2019qoh} for more detailed explanations as well as how to obtain a viable cosmology in this class of models.

Fig.~\ref{fig:theoryspace} presents the theoretically-motivated region of parameter space for this heavy, high-quality axion model, as a function of the axion mass $m_a$ and the axion decay constant $f_a\equiv f_G/(4\pi^2)$. Since our mechanism makes the axion only heavier, it can not populate the region labeled ``QCD Axion'' where it would be lighter than the QCD axion. In the region labeled ``$f_a<\Lambda_{\rm QCD'}$'', the axion EFT breaks down because $m_a>f_a$ in that region. In the region labeled ``PQ Quality Problem,'' the axion suffers the Quality Problem discussed in Eq.~\eqref{eq.quality} due to operators with $N \geq 6$. Finally, in the region in the bottom right denoted ``$H'$ Quality Problem'', the mirror VEV $\langle H' \rangle$ spoils the Strong CP solution via,
\begin{align}\label{eq.higgsvevqual}
\frac{\alpha_3}{8\pi}\left(\frac{H^\dagger H}{M_{\rm pl}^2}G\tilde{G}+\frac{H'^\dagger H'}{M_{\rm pl}^2}G'\tilde{G'}\right).
\end{align}
In particular, to avoid the Quality Problem from Eq.~\eqref{eq.higgsvevqual}, we require $\langle H'\rangle < 10^{14}$~GeV, so there is only a maximal amount by which the axion can be made heavier in this scenario.\footnote{While a portal coupling $\lambda |H|^2 |H'|^2$ can be present, its primary effect would be to make the SM Higgs very heavy unless $\lambda$ is very small.  We view this generically large contribution to the Higgs mass as another form of the hierarchy problem for the SM Higgs which we do not try to address in this work. }

By inspection, Fig.~\ref{fig:theoryspace} encourages us to focus on heavy axions in the keV-TeV mass range with $f_G$ between $10-10^9$ GeV. A natural question that emerges is how much of the open theoretical parameter space can be covered by existing and upcoming experiments. To discuss this, we detail a phenomenological discussion of the heavy axion properties next.

\subsection{Heavy Axion EFT, Mixing and Lifetime}\label{subsec:EFT}
A robust consequence of the above mentioned class of models is the
defining $G\tilde{G}$ coupling of the axion. For this purpose, we consider an effective Lagrangian,
\begin{equation}\label{eq:alpeft}
\frac{a}{8\pi f_a}\left(c_3 \alpha_3 G\Tilde{G}+c_2\alpha_2 W\Tilde{W}+c_1\alpha_1 B\Tilde{B}\right),
\end{equation}
with $\alpha_i=g_i^2/(4\pi)$ given in terms of SM gauge couplings, and $\alpha_1=5/3\alpha_Y$, in terms of the hypercharge gauge coupling. 
To illustrate the significance of a non-zero $c_3$, we will focus on two scenarios which are complementary to the case of photon or electroweak-dominance, $c_2,c_1\gg c_3$, frequently assumed in the literature due to its testability. In more detail, we will focus on the cases of,
\begin{itemize}
\item {\bf Gluon dominance}: $c_3=1, c_1, c_2=0$;
\item {\bf Codominance}, $c_1=c_2=c_3$.
\end{itemize}
Both of the above cases are motivated from the generic Axion considerations, as well as UV considerations. In fact, these two cases match well respectively to the KSVZ~\cite{Kim:1979if,Shifman:1979if} and DFSZ~\cite{Dine:1981rt,Zhitnitsky:1980tq} scenario of the minimal axion theory.

These choices have an important effect on the phenomenology of such heavy axions, since for $m_a\gtrsim 1$~GeV, the axions predominantly decay into hadronic final states, as opposed to diphoton final states on which a significant number of searches rely.\footnote{For a recent study in the $aF\tilde F$ dominance at DUNE ND, see Ref.~\cite{Brdar:2020dpr}.} Consequently, interesting parts of the axion parameter space open up, as we will see below. Simultaneously, a non-negligible $c_3$ gives rise to important axion production channels at the LHC and various proton beam dump experiments.

Below the scale of electroweak symmetry breaking, the EFT in Eq.~\eqref{eq:alpeft} gives rise to an axion photon coupling,
\begin{align}
\frac{a}{8\pi f_a}c_\gamma\alpha_{\rm EM}F\tilde{F},
\end{align}
with \cite{Georgi:1986df,Bauer:2017ris,Aloni:2018vki}
\begin{align}
c_\gamma =& c_2+\frac{5}{3} c_1~~\text{for}~~m_a\gg \Lambda_{\rm QCD}\nonumber\\
c_\gamma =& c_2+\frac{5}{3} c_1 + c_3\left(-1.92+\frac{1}{3}\frac{m_a^2}{m_a^2-m_\pi^2}+\frac{8}{9}\frac{m_a^2-\frac{4}{9}m_\pi^2}{m_a^2-m_\eta^2}+\frac{7}{9}\frac{m_a^2-\frac{16}{9}m_\pi^2}{m_a^2-m_{\eta'}^2}\right)~~\text{for}~~m_a\lesssim \Lambda_{\rm QCD}\label{eq.cgalow}.
\end{align}
To obtain $c_\gamma$ for $m_a\lesssim \Lambda_{\rm QCD}$, we have assumed the $\eta-\eta^\prime$ mixing angle $\sin\theta_{\eta\eta^\prime}=-1/3$ as in \cite{Aloni:2018vki} while noting the significant uncertainty $\theta_{\eta\eta^\prime}\simeq-(10^\circ-20^\circ)$ (see Sec.~15 of Ref.~\cite{Zyla:2020zbs} and references therein). Some results for more general mixing angles can be found in Refs.~\cite{Ertas:2020xcc,Gori:2020xvq}. The factor of $1.92$ in Eq.~\eqref{eq.cgalow} can be obtained after including higher order corrections \cite{diCortona:2015ldu} on the leading order contribution due to quark masses $\frac{2}{3}\frac{4m_d+m_u}{m_u+m_d}\approx 2$. Importantly, we see that even in the absence of the a tree-level $c_1,c_2$, the anomaly and axion-pseudoscalar meson mixing introduce a significant axion-photon coupling below $\Lambda_{\rm QCD}$.

\begin{figure}
\begin{center}
\includegraphics[width=0.6\linewidth]{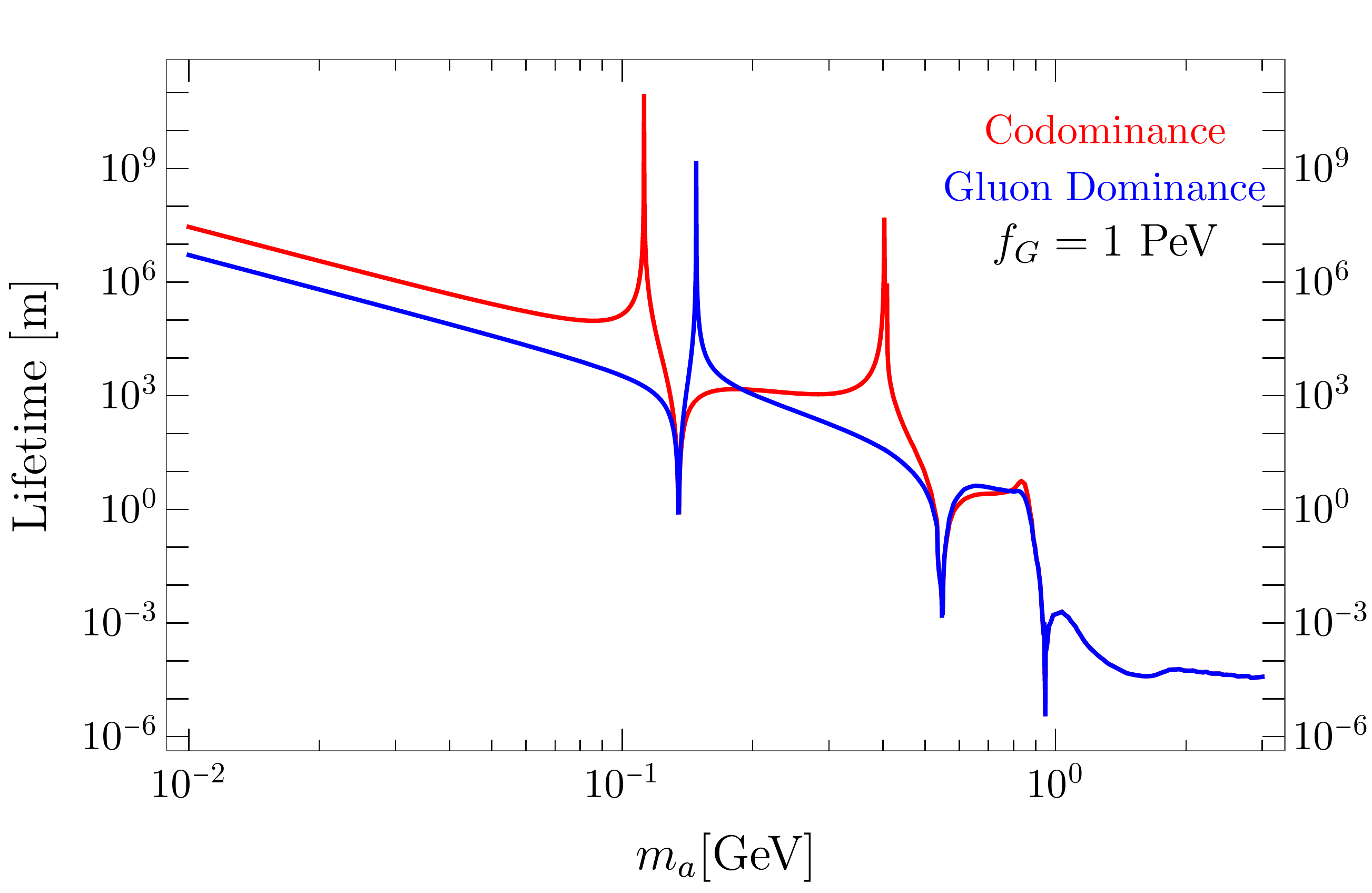}
\caption{Lifetime of the heavy axion for a decay constant $f_G=1~\rm{PeV}$ as a function of its mass $m_a$. The blue (red) line assumes the Gluon Dominance (Codominance) scenario discussed in the text.
}
\label{fig:lifetime}
\end{center}
\end{figure}

The phenomenology will be largely dictated by the lifetime of the axion. For $m_a<3m_\pi$, the axion decays exclusively in diphoton final states with a width,
\begin{align}
\Gamma_{\gamma\gamma}=\frac{\alpha_{\rm EM}^2c_\gamma^2}{256\pi^3}\frac{m_a^3}{f_a^2},
\end{align}
whereas above that threshold the hadronic decay modes open up \cite{Aloni:2018vki} and quickly become dominant except regions near resonant mixing.
We show the resulting lifetime of the axion in Fig.~\ref{fig:lifetime} which is used in the following to determine the reach of the DUNE ND.
The blue (red) line in Fig.~\ref{fig:lifetime} assumes the Gluon Dominance (Codominance) scenario, and both lines assume $f_G = 1$ PeV. This lifetime is proportional to $f_G^2$. We note that the two scenarios are nearly identical for $m_a \gtrsim 1$ GeV but %have significant differences 
below $1$ GeV the above distinctions are quite important. For $m_a$ below 100~MeV, we can see that in the Codominance scenario, the axion has a larger lifetime as the corresponding $c_\gamma$ is smaller for the particular choices of $c_i=1$. Between 100~MeV and 1~GeV, two effects are important. One is the near resonance mixing with the SM mesons, which determines the dips in this lifetime plot. The other one is the cancellation between different meson mixings and the direct contributions to $c_\gamma$ in Eq.~\ref{eq.cgalow} for which we get peaks in the lifetime. We also note here, the mixing expansion in this equation at the meson pole regime should be regulated by the unitarity of the mixing matrix, which we neglected here in the equations but implemented effectively in the next section in our numerical computation. 

Before moving on, we note that due to the $Z_2$ symmetry, the specific model described in Ref.~\citep{Hook:2019qoh} and above, has a massless mirror photon.  While the axion can decay into a pair of mirror photons, for our phenomenlogical analysis below we will ignore this effect, motivated by the following reason.  The mirror photon does not play an essential role in our set up and can be removed from the spectra if we do not copy the SM $U(1)_Y$ into the mirror sector. Instead, we can start with a common $U(1)_X$ under which both SM and mirror sectors are charged. As the mirror Higgs gets a VEV, the breaking $SU(2)_W^\prime\times U(1)_X\rightarrow U(1)_Y$ takes place without giving rise to a mirror photon. Since the flavor structure of both the SM and the mirror sector are the same, the effects of differential RG running of $\bar{\theta}$ between the two sectors are still suppressed.  %Second,  even without this UV story, the effects of the mirror photon is negligible in the Gluon dominance scenario. This is because, there the axion-photon coupling is inherited from the SM psuedoscalar-photon couplings and the axion-SM pseudoscalar mixings.  But, this effect is negligible for the mirror sector since the analogs of the SM pseudoscalars are very heavy $\gg m_a$, and consequently axion-mirror photon coupling is not induced significantly in the Gluon dominance scenario.  

%On the other hand, in the Codominance scenario, the axion can indeed pick up a mirror photon coupling due to non-zero $c_1,c_2$. Consequently, axion branching ratio to mirror photons is $\mathcal{O}(1)$ comparable to that into SM photons. However as we will see,  for the novel, large-lifetime parameter space probed by the DUNE ND, its reach sensitivity to axion decay constant scales as $f_a^4$ and hence, the $\mathcal{O}(1)$ change in the branching ratio only has a small $\mathcal{O}(1)^{1/4}$ effect on the reach of $f_a$.  Because of this reason,  and also to be least model-dependent from an IR EFT point of view, we will ignore the axion decays into mirror photon.

\section{Simulation Details: Axion Production and DUNE Near Detector}\label{sec.axionsim}
This section details the simulations we perform and how we determine the DUNE experimental sensitivity to heavy axions. In Section~\ref{sec:ProductionDetails} we explain the approach we employ to calculate the axion production, both from meson mixing and from gluon-gluon fusion. Section~\ref{subsec:DetectorSpecifics} explains how we include the DUNE ND complex in these simulations, including both the liquid argon near detector and the gaseous argon multi-purpose detector.

\subsection{Axion Production Details}\label{sec:ProductionDetails}
\textbf{Meson Mixing:} To determine the axion production due to meson mixing, we simulate the Long Baseline Neutrino Facility (LBNF) beam as a 120 GeV proton beam colliding with a fixed target using \texttt{Pythia8} with the ``\texttt{SoftQCD:all = on}'' option. %\footnote{We have considered proton-neutron collisions for concreteness. Proton-proton collisions give similar results.}
For all of our simulations, we assume the total number of protons-on-target (POT) is $N_{\rm POT} = 1.47 \times 10^{22}$ over the course of 10 years.\footnote{This assumes ten years of operation at the nominal rate of $1.47 \times 10^{21}$ POT/yr. The DUNE collaboration plans on upgrading its beam to a larger number of POT/yr during its operation, so our estimations should correspond to \textit{at most} ten years of data collection.} We find that approximately $2.89$ $\pi^0$, $0.33$ $\eta$, and $0.03$ $\eta^\prime$ are produced per POT at this beam energy.

As alluded to in Eq.~\eqref{eq.cgalow}, axions mix with the SM pseudoscalar mesons through the $G\tilde{G}$ coupling. Here we summarize the mixing angles \cite{Bauer:2017ris,Aloni:2018vki,Ertas:2020xcc},
\begin{align}\label{eq.api}
\pi =& \pi_{\rm phys} + \theta_{a\pi}a_{\rm phys} +\cdots \approx \pi_{\rm phys}+\frac{1}{6}\frac{f_\pi}{f_a}\frac{m_a^2}{m_a^2-m_\pi^2}a_{\rm phys}+\cdots,\\
\label{eq.aeta}
\eta =& \eta_{\rm phys} + \theta_{a\eta}a_{\rm phys} +\cdots \approx \eta_{\rm phys}+\frac{1}{\sqrt{6}}\frac{f_\pi}{f_a}\left(\frac{m_a^2-\frac{4}{9}m_\pi^2}{m_a^2-m_\eta^2}\right)a_{\rm phys}+\cdots,\\
\label{eq.aetap}
\eta^\prime =& \eta^\prime_{\rm phys} + \theta_{a\eta^{\prime}}a_{\rm phys} +\cdots\approx \eta_{\rm phys}^\prime+\frac{1}{2\sqrt{3}}\frac{f_\pi}{f_a}\left(\frac{m_a^2-\frac{16}{9}m_\pi^2}{m_a^2-m_{\eta^{\prime}}^2}\right)a_{\rm phys}+\cdots,
\end{align}
where $f_{\pi}\approx 93~\rm MeV$. 
In these equations, the ellipses contain $\pi-\eta$ and $\pi-\eta^\prime$ mixing terms, which subdominantly contribute to ALP production considered below. 

\begin{figure}
\begin{center}
\includegraphics[width=0.6\linewidth]{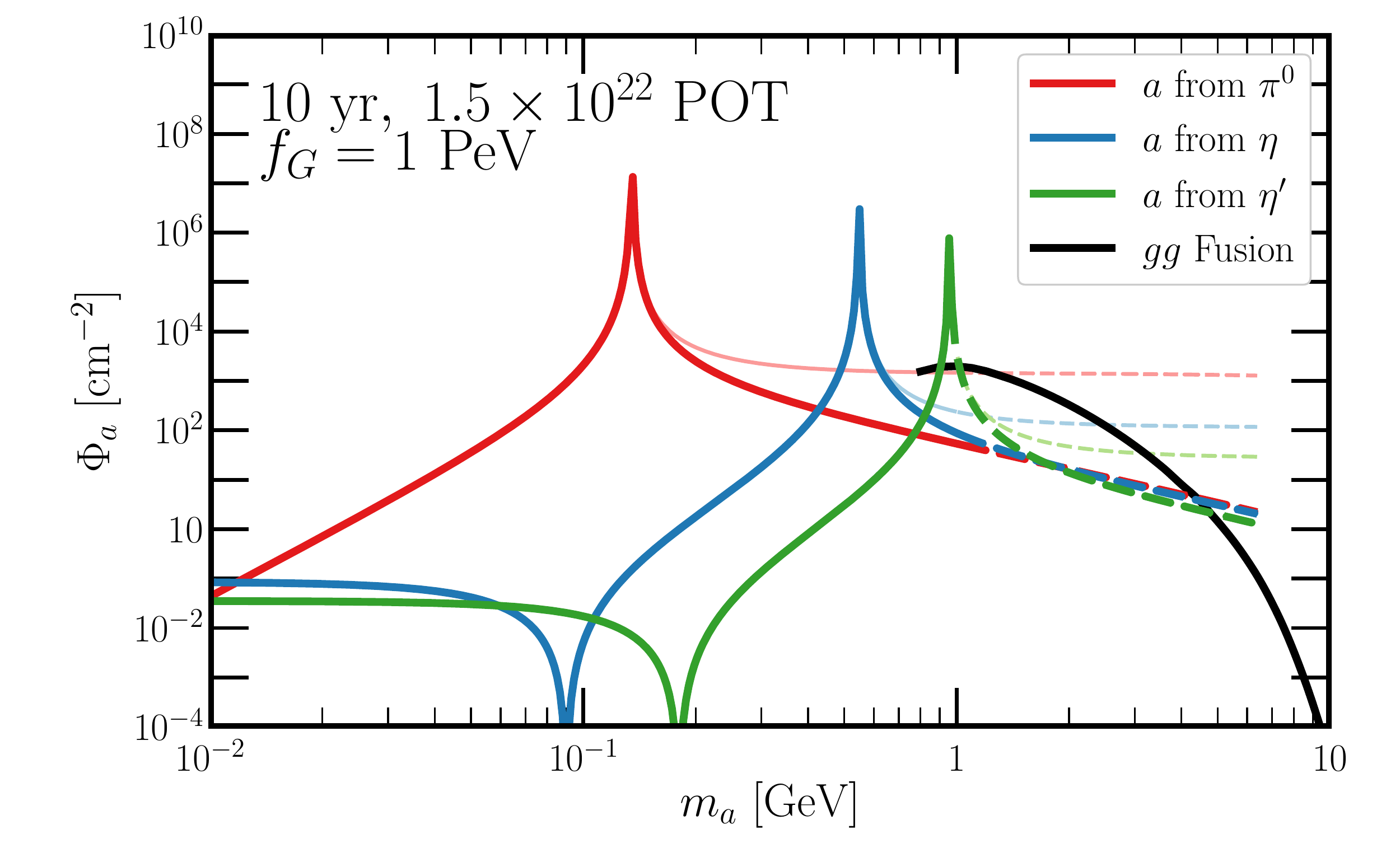}
\caption{Expected flux at the DUNE Near Detector Hall of heavy axions with mass $m_a$ produced via meson mixing with $\pi^0$ (red), $\eta$ (blue), and $\eta^\prime$ (green), along with gluon-gluon fusion (black), all assuming $f_{G} = 1$ PeV (or, equivalently $g_{agg} = 1$ PeV$^{-1}$). The flux scales with $f_{G}^{-2}$. We dash meson-mixing flux lines above $m_a = 1$ GeV where gluon-gluon fusion processes are more relevant. For reference, we also show the the flux from meson mixing one would get without the mass correction (taking the function $f(m_{\mathrm {meson}},m_a)=1$) in thin lines.
\label{fig:DUNEFlux}}
\end{center}
\end{figure}
Using Eqs.~\eqref{eq.api}, \eqref{eq.aeta} and \eqref{eq.aetap}, the number of axions produced from ALP-meson mixing is obtained as \footnote{We ignore possible interference effects between different meson-mixing modes in this approximation.},
\begin{align}\label{eq:Na}
N_{\rm axions}=N_{\rm POT}\times\left[2.89|\theta_{a\pi}|^2 f(m_\pi, m_a) +0.33|\theta_{a\eta}|^2 f(m_\eta, m_a)+0.03|\theta_{a\eta^\prime}|^2 f(m_{\eta^\prime}, m_a)\right],
\end{align}
where
$$
f(m_{\mathrm{meson}},m_a)=
\begin{cases}
\left(\frac {m_a} {m_{\mathrm{meson}}}\right)^{-1.6}    & \quad \text{if }  m_a> m_{\mathrm{meson}}\\
    1  & \quad \text{if } m_a\leq m_{\mathrm{meson}}.
\end{cases}
$$
The above function $f(m_{\mathrm{meson}},m_a)$ models the QCD production rate of mesons which decreases as one increases the meson mass, rooting from both the running strong coupling as well as the parton evolution. The power of $-1.6$ comes from fitting the $\pi^0$, $\eta$, and $\eta^\prime$ meson production rate as a function of their masses. Furthermore, we conservatively bound the function value by unity, by neglecting the possible enhancement of rate beyond the mixing calculation in the regime of $m_a\leq m_{\mathrm{meson}}$. Note that to our knowledge we are the first to take this further step to model the kinematic effect of masses in this regime of the axion production rate. To show the difference, we show the flux in Fig.~\ref{fig:DUNEFlux} with and without this mass effect taken into account in thin and thick curves. 

The axion flux with axion mass below 1~GeV at the DUNE ND then depends on Eq.~\eqref{eq:Na}, the ND cross-sectional area, and the acceptance fraction. 
In detail, for each simulated $\pi^0$, $\eta$, and $\eta^\prime$, we convert them into an axion with a weight according to the mixing angle. This conversion process keeps the energy of the SM meson the same, rescaling the magnitude of the three-momentum while maintaining their direction.
The acceptance fraction is defined as the fraction of produced axions that are traveling in the direction of the DUNE ND upon production, folding in the production angular dependence in the beam dump environment. We discuss the details of the DUNE ND in our simulations in Section~\ref{subsec:DetectorSpecifics}, and we note here that the acceptance fraction for the different meson-mixing production mechanisms is $\mathcal{O}(10^{-2})$.

Fig.~\ref{fig:DUNEFlux} displays the expected flux at the DUNE ND for axion production from meson mixing as a function of the mass $m_a$. We show the separate contributions from $\pi^0$, $\eta$, and $\eta^\prime$ as different colors, which allows us to see the different axion-meson mixing dominating for different regions of $m_a$. We dash the contributions for $m_a \gtrsim 1$ GeV, where we expect gluon-gluon fusion to serve as a better description of axion production in this environment. This flux is shown for $f_{G} = 1$ PeV and scales with $f_{G}^{-2}$.

\textbf{Gluon-Gluon Fusion:} Above $\mathcal{O}$(GeV) masses, the direct production mode from gluon-gluon fusion could potentially dominate the contribution to the heavy axion flux at beam dump facilities. There, given the momentum exchange is above the GeV scale, the parton distribution function description is valid. The operator $\propto aG\tilde G$ determines the production rate.

\begin{figure}
\begin{center}
\includegraphics[width=0.48\linewidth]{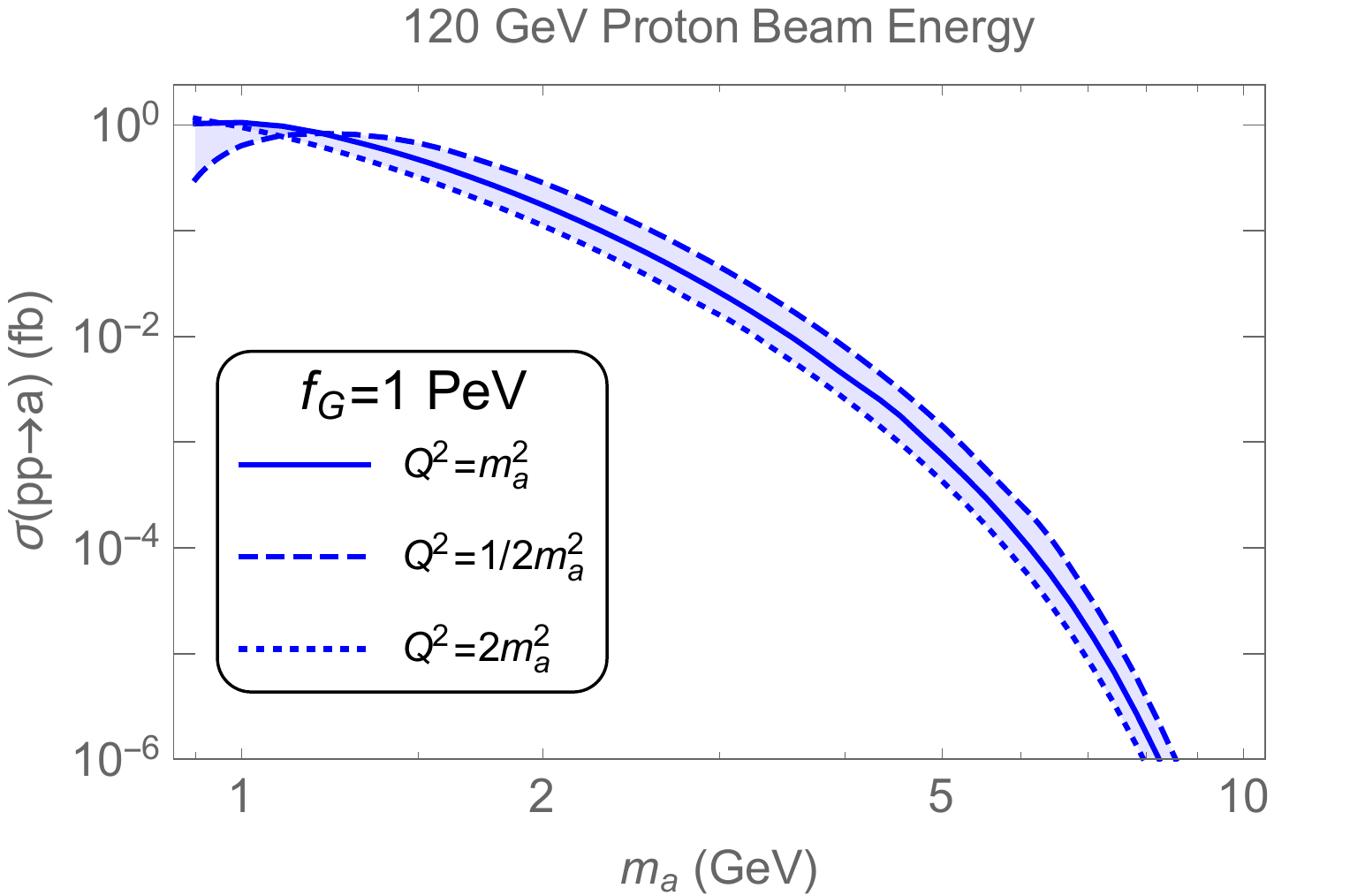}
\includegraphics[width=0.48\linewidth]{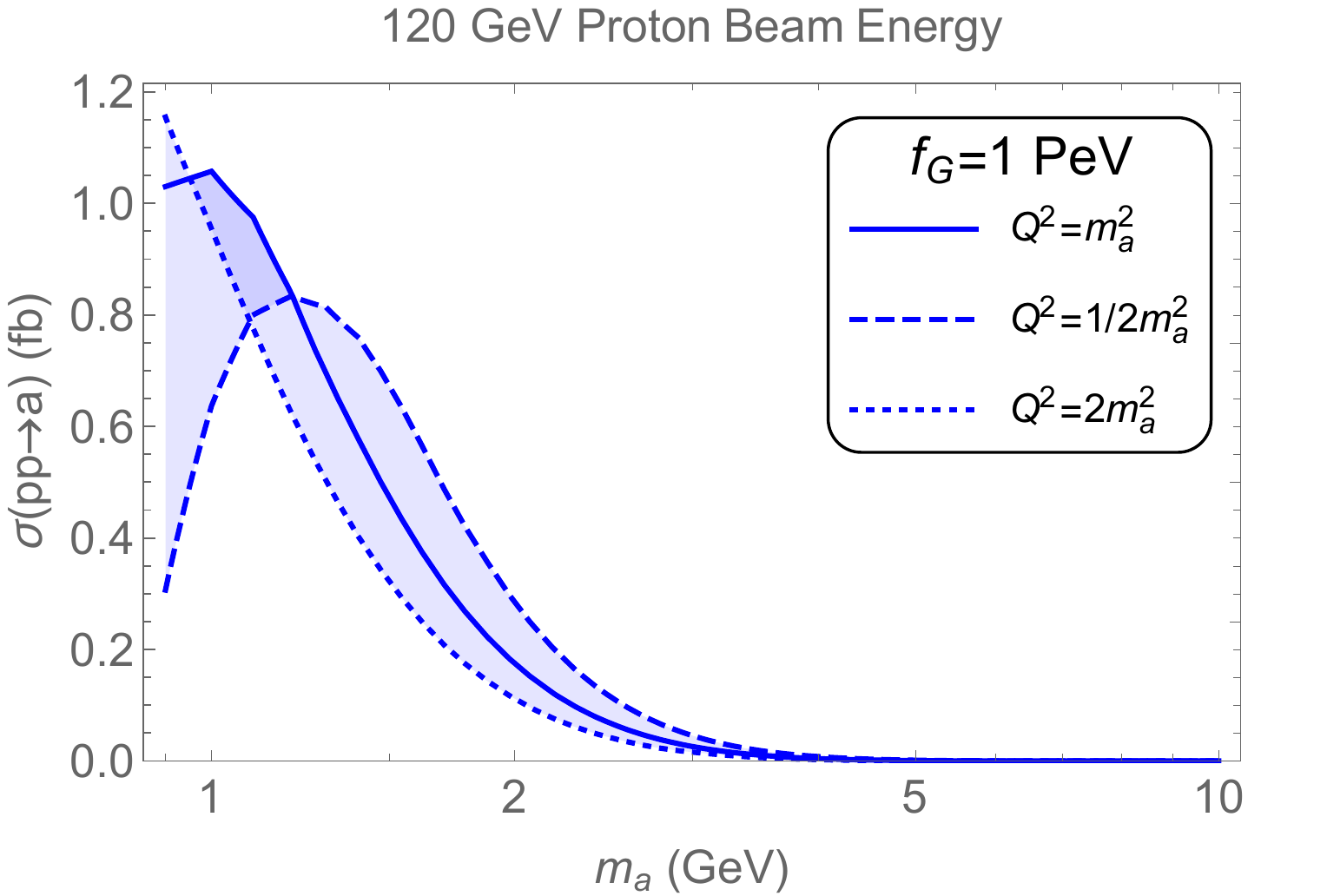}
\caption{The gluon-gluon fusion production rates for heavy axions at the LBNF beam for the DUNE experiment. We include the scale uncertainty of the production from both the PDF side and running of $\alpha_S(Q^2)$, with scale choices of $Q^2=\left\lbrace1,0.5,2\right\rbrace m_a^2$ shown in solid, dashed and dotted lines, respectively.
\label{fig:ggF}}
\end{center}
\end{figure}

We evaluate the production cross section convoluted with the leading order parton distribution function {\tt NNPDF}~\cite{Ball:2014uwa,Hartland:2012ia} with our calculation available at \href{https://gitlab.com/ZhenLiuPhys/alpdune}{\tt this link}, following:
\beq
\sigma(pp\to a)=\frac {\alpha^2_s(\mu_R^2) m_a^2} {256\pi f_a^2 s} \int^1_{m_a^2/s} dx \frac 1 x f_g(x, \mu_F^2) f_g(\frac {m_a^2} {x s}, \mu_F^2) ,
\eeq
where $\mu_F$ and $\mu_R$ are the factorization and renormalization scale, and $s$ is the center of mass energy, approximately $(15~\gev)^2$.
The factorization and renormalization scale is set at the heavy axion mass $\mu_R^2=\mu_F^2=m_a^2$. In Fig.~\ref{fig:ggF}(left), we show the inclusive production rate for a decay constant $f_G=1~$PeV. We used the 1-loop renormalization-group-equation running of the strong coupling constant $\alpha_S(Q^2)$ embedded in {\tt NNPDF}. In Fig.~\ref{fig:ggF}(right), we show the same rate in a linear scale, effectively zooming in around $m_a\approx 1$ GeV. We also show how the cross section varies when we consider factorization scales between $m_a^2$, $1/2 m_a^2$ (dashed lines) and $2m_a^2$ (dotted). In both panels, we see the scale uncertainty is sizable, especially below the GeV scale. In fact, this regime is where PDFs have large uncertainties and scheme dependence. We observe with our current NNPDF choice the cross section only starts to dominate when $m_a \gtrsim$ 0.9 GeV, and so our results are not subject to the large uncertainty in the low mass regime. Another interesting phenomena is the cross-over of different scale choices, this is a result of the PDF evolution and the $\alpha_S^2 (Q^2)$ running from the production cross section.

\begin{figure}
\begin{center}
\includegraphics[width=0.6\linewidth]{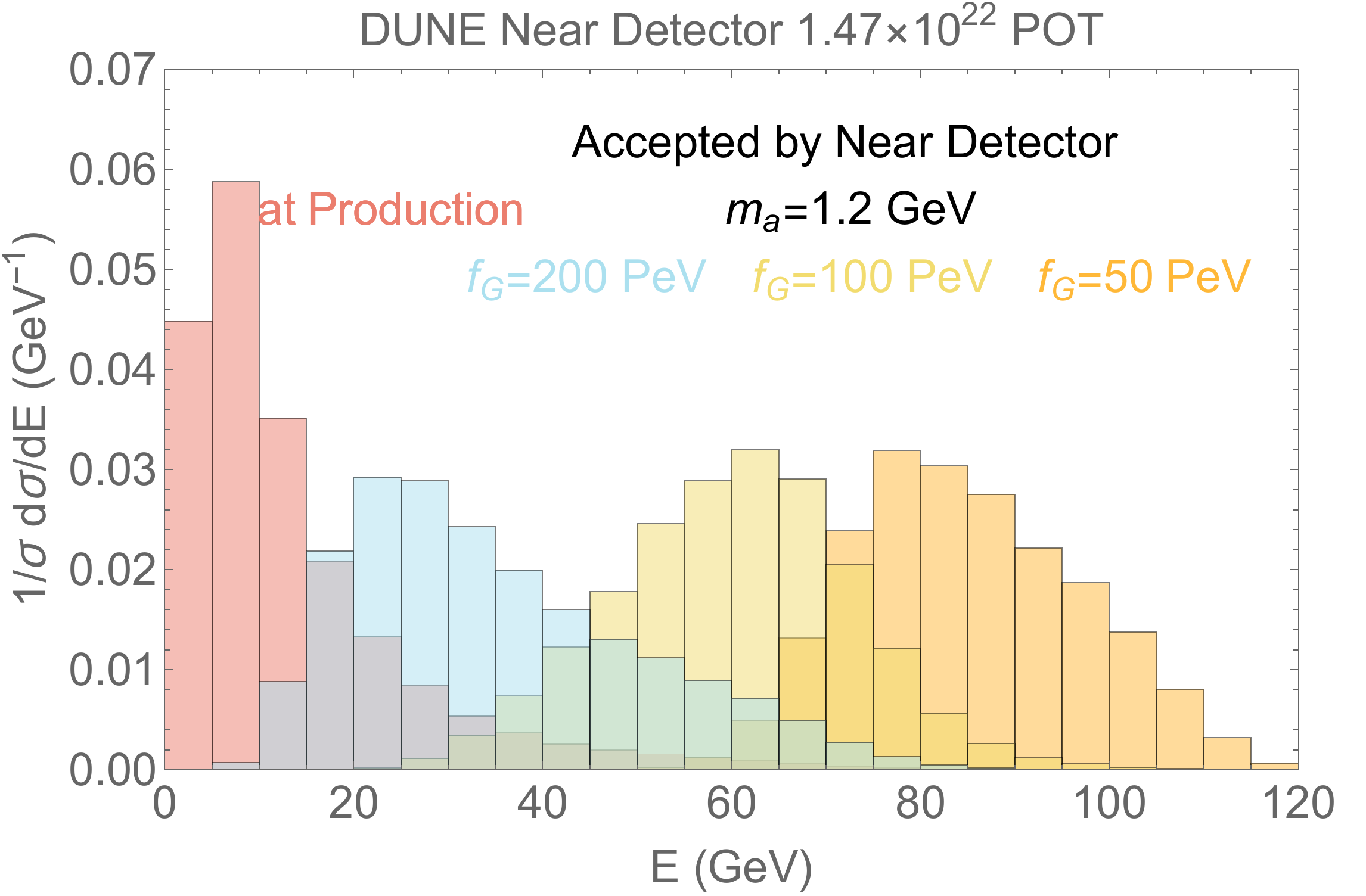}
\caption{The energy distributions for a heavy axion from gluon-gluon fusion at production and after acceptance by the DUNE Near Detector. We show the distributions all for $m_a = 1.2$ GeV and for the total sample of all produced $a$ (red), as well as for those that reach the Near Detector before decaying for  $f_G= 50$ (orange), $100$ (yellow), and $200$ (cyan) PeV.
\label{fig:ggFdist}}
\end{center}
\end{figure}

In Fig.~\ref{fig:ggFdist}, we show the normalized energy distribution from the gluon-gluon fusion production for a benchmark heavy axion with a mass of 1.2~GeV. The red curves show the distribution at production, which is universal for all axion decay constants. For $m_a\ll \sqrt{s}$ , where $\sqrt {s}\simeq 15$~GeV for the LBNF-DUNE beam under consideration here, the distribution from gluon-gluon fusion are similar. In the same plot, we also show the differential distribution in orange, yellow and cyan for the axion accepted by the DUNE ND (accounting for decays of axions en route) for axion decay constant $f_G$ of 50, 100, and 200 PeV, respectively. The shift in distribution is mainly driven by the necessary boost for a heavy axion to arrive at the DUNE ND before decaying. More boost is needed for a smaller decay constant, and hence the distribution shifts towards higher energies. We discuss the detection considerations and details in Section~\ref{sec:ExpSignatures}.

A few other axion production modes could be significant. These include bremsstrahlung effects from the proton-proton collinear emission with resummation. The result will depend on how one treats the finite mass effect from the axion and its derivative coupling. Similarly, there can be collinear emissions from the quarks and gluons in the collision, involving model-dependent axion-quark couplings. Through the axion coupling to quarks, possible flavor-changing decays from mesons will also contribute to axion production~\cite{MartinCamalich:2020dfe}.
Last but not least, the proton-proton collision at beam dumps will create secondary collisions from the remnants of the first collision, enlarging the number of mesons produced and hence enriching the flux for heavy axions below $\mathcal{O}\mathrm{(GeV)}$. All these effects could help improve the axion flux and, therefore, the DUNE ND sensitivities to heavy axions. We leave a detailed analysis of these different contributions with more model dependence for future studies.

% \textbf {Proton Bremstrahlung:}

% \textbf {Other production modes:}
% Meson decays~\cite{MartinCamalich:2020dfe} from flavor changing effects.

\subsection{DUNE Near Detector Complex Details}\label{subsec:DetectorSpecifics}
We are interested in signatures of axion decay in the DUNE ND Complex. Specifically, we consider such signatures inside the liquid argon time-projection-chamber detector ArgonCube and the gaseous argon time-projection-chamber Multi-Purpose Detector (MPD). ArgonCube is situated at a distance of 574 m from the DUNE proton target and has a total active volume of $7$ m wide, $3$ m high, and $5$ m long. Fiducialization reduces this to a fiducial mass of roughly $67.2$ t~\cite{Abi:2020evt,Abi:2020kei}. The MPD is situated directly downstream of ArgonCube (designed to be a spectrometer of muons and other particles that do not stop in ArgonCube) with a cylindrical volume that is roughly 5 m in diameter and 5 m in height. This corresponds to an active mass of $1$ ton. The MPD is situated inside an electromagnetic calorimeter and a magnetic field, allowing for precision measurement (and charge and particle identification) of the particles traveling through its fiducial volume. Fig.~\ref{fig:NDDrawing} provides a schematic drawing of the DUNE target and Near Detector Complex (note that many elements are removed from this figure for simplicity, including the magnetic focusing horns and a significant amount of earth between the decay volume near the target and the detector hall).

Given the dimensions of the detectors and the DUNE target/detector distance, we find that $\mathcal{O}(10^{-2})$ of the axions produced via meson mixing will travel in the direction of the near detector complex. % and have the capability of decaying within.
We include axion decays inside both the liquid and gaseous detectors in our simulations, corresponding to a total decay length of roughly $10$ m. In Section~\ref{sec:ExpSignatures}, we discuss the various experimental signatures of this heavy axion decay and how we can reduce associated backgrounds in the two detectors.

\section{Experimental Signatures of Heavy Axion Decay}\label{sec:ExpSignatures}
For the ALP masses we expect to be sensitive to at the DUNE ND, two classes of $a$ decays are of interest: $a \to \gamma\gamma$ and $a\to$ hadrons. For $m_a \lesssim 1$ GeV, the latter ($a \to$ hadrons) consists mostly of $a \to \pi\pi\pi$ and $\pi\pi\gamma$. Here, we highlight the characteristics of these respective signals in the DUNE ND complex focusing on both ArgonCube and MPD.

In Table~\ref{tab:SignalBackground} we list some defining characteristics of the two signals, $a \to \gamma\gamma$ and $a\to$ hadrons, both in ArgonCube and MPD. We also list the types of neutrino-scattering backgrounds that contribute to these searches, and some properties of the backgrounds that allow for separating our signal from these events.
\begin{table}
\begin{center}
\caption{Signals of ALP decay $a \to \gamma \gamma$ and $a\to$ hadrons in the liquid argon near detector (ArgonCube) and the gaseous argon detector (MPD). We also list the dominant source of backgrounds in the detectors for each of these searches, and some properties that distinguish between the signals and backgrounds.
\label{tab:SignalBackground}}
\begin{tabular}{|c||c|c||c|c|}\hline
\multirow{2}{*}{\textbf{Signature}} & \multicolumn{2}{c||}{\textbf{Liquid Argon ArgonCube}} & \multicolumn{2}{c|}{\textbf{Gaseous Argon MPD}} \\ \cline{2-5}
& \textbf{Signal} & \textbf{Background} & \textbf{Signal} & \textbf{Background} \\ \hline
\multirow{4}{*}{$a \to \gamma \gamma$} & Invariant Mass & NC$\pi^0$ & Invariant Mass & NC$\pi^0$ \\
& $\gamma\gamma$ Direction & Nearly-Isotropic & $\gamma\gamma$ Direction & Nearly-Isotropic \\
& High-Energy & Low-Energy & High-Energy & Low-Energy \\
& & & & Low-energy recoils \\ \hline
\multirow{5}{*}{$a \to$ hadrons} & Invariant Mass & CC$1\mu2\pi$ & Invariant Mass & CC$1\mu2\pi$ \\
& Opening angle & DIS & Opening angle & DIS \\
& High-energy & Low-energy & High-Energy & Low-Energy \\
& $gg$ Direction & Nearly-Isotropic & $gg$ Direction & Nearly-isotropic \\
& & & & Low-energy recoils \\ \hline
\end{tabular}
\end{center}
\end{table}
For both signatures, and in both detectors, the $a$ decay will be very forward---the large boost factor of $a$ and the decay kinematics require this. Meanwhile, background events, such as those from neutral-current (NC) $\pi^0$ production (and subsequent $\pi^0 \to \gamma\gamma$ decay) will be more isotropic, and may also have some measurable nuclear recoil that would not be present in the signal. In our signal $a \to \gamma\gamma$, since it is a fully visible decay, the invariant mass $m_{\gamma\gamma}^2 = m_a^2$, whereas the background events should reconstruct $m_{\pi^0}^2$ in this case. Additionally, the $\pi^0$ are being produced by neutrino scattering and will have at most $E_{\pi^0} \lesssim 5$ GeV (a conservative estimate). Lastly, the large boost of the $a$ will result in small opening angles in the $\gamma\gamma$ final state, whereas the (less-boosted) $\pi^0$ from NC production will have larger opening angles.

\begin{figure}
\begin{center}
\includegraphics[width=0.485\linewidth]{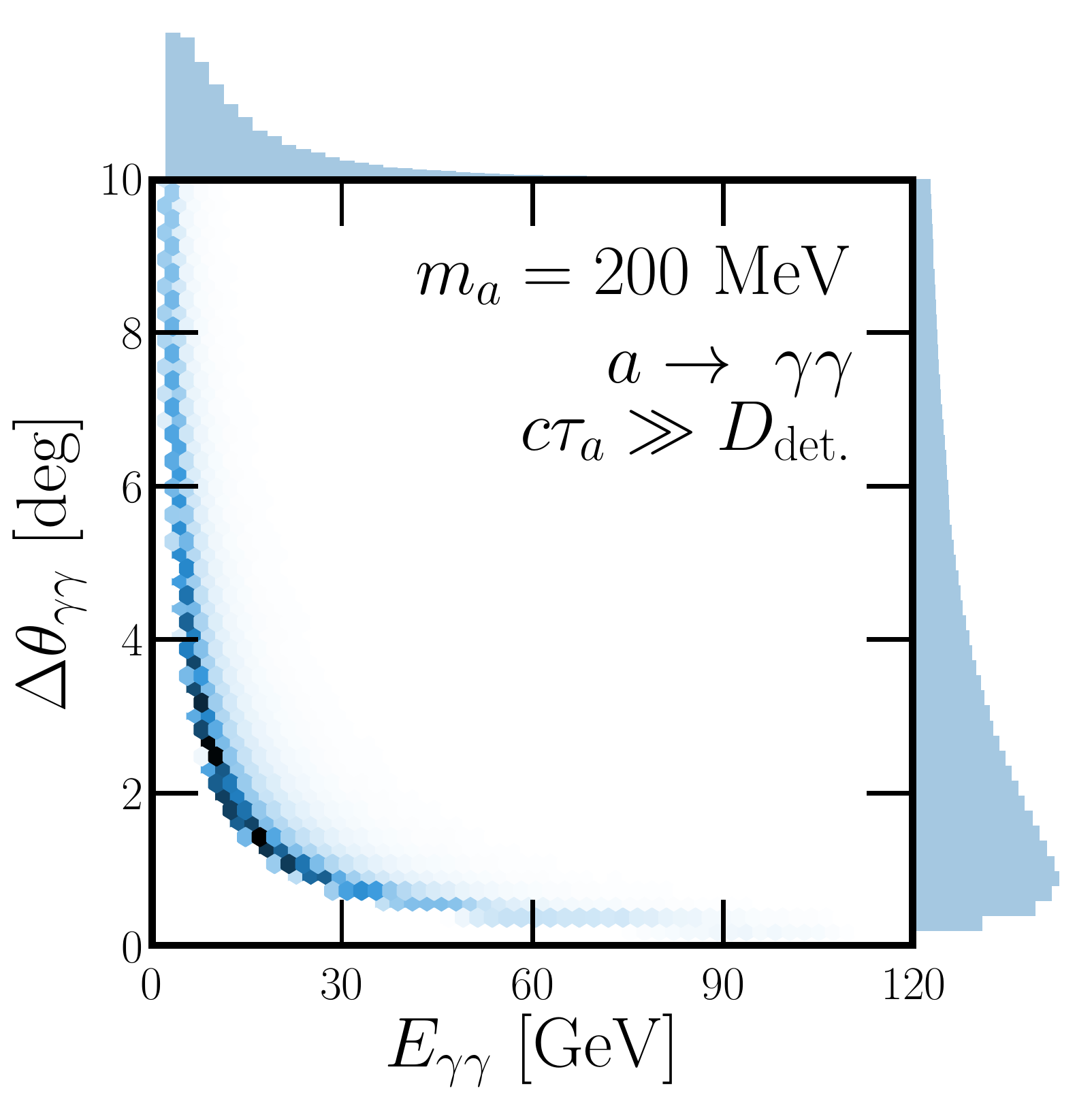}
\includegraphics[width=0.485\linewidth]{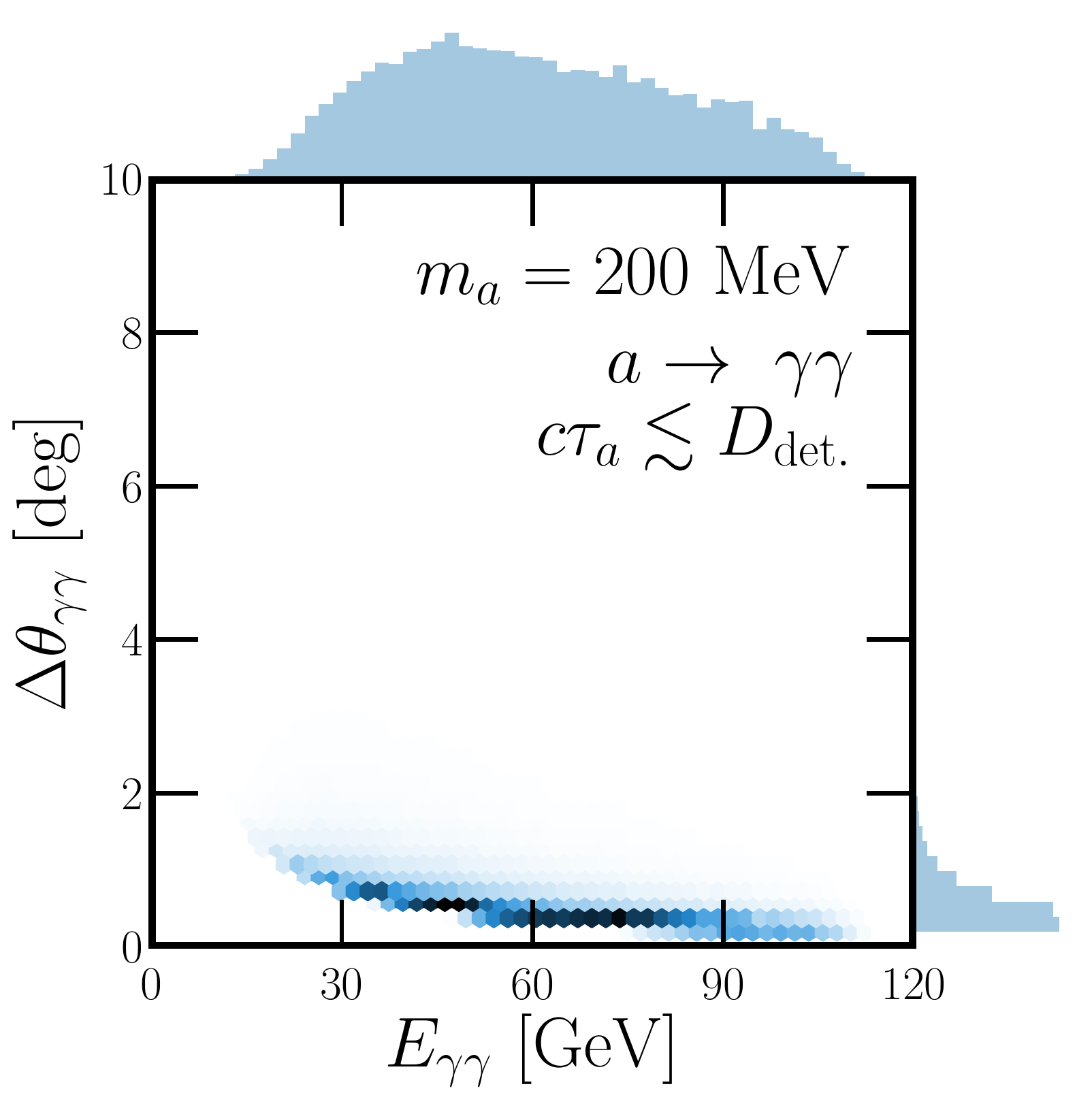}
\caption{Signal distributions for an axion with $m_a = 200$ MeV, produced via meson mixing (with $\pi^0$) decaying in the DUNE Near Detectors, with respect to the total diphoton energy $E_{\gamma\gamma}$ and opening angle between the photons $\Delta\theta_{\gamma\gamma}$ for the decay channel $a \to \gamma\gamma$. The left panel assumes $a$ is long-lived relative to the distance between the DUNE target and detector, whereas the right panel assumes it is short-lived. One-dimensional distributions for each of these observables are shown on top of/to the right of the two-dimensional distribution panels.
\label{fig:2DDist_MesMix}}
\end{center}
\end{figure}

We illustrate a subset of these distinctions in Fig.~\ref{fig:2DDist_MesMix}, where we show two different signal event distributions for a $200$ MeV axion produced via meson mixing and decaying in the DUNE ND with the signal $a\to \gamma\gamma$. The main panels in Fig.~\ref{fig:2DDist_MesMix} (left) and (right) display the distribution of these signal events as a function of the total diphoton energy $E_{\gamma\gamma}$ as well as the angle between the two outgoing photons in the lab frame, $\Delta\theta_{\gamma\gamma}$. On the top (right) of the main panels, one-dimensional histograms display the distributions of $E_{\gamma\gamma}$ ($\Delta\theta_{\gamma\gamma}$) independently.
Both panels of Fig.~\ref{fig:2DDist_MesMix} correspond to an axion with $m_a = 200$ MeV produced from meson mixing, predominantly with $\pi^0$. The distinction between these two panels is in the lifetime of $a$, i.e., whether it is long-lived (left) or short-lived (right), relative to the target-detector distance of 574 m. In the left panel, we assume $a$ is long lived and $c\tau \gg$ 574 m. Here, the probability of a given $a$ to decay in the detector is proportional to $(\gamma c\tau)^{-1}$ which scales as $m_a/E_a$ and favors lower-energy $a$ from the production distribution. In contrast, the right panel assumes $a$ is short-lived, $c\tau \ll$ 574 m. In this scenario, only the high-energy $a$ have $\gamma$ high enough that their time-dilated lifetime is on the order of the target-detector distance and can survive that journey. This results in high-energy $a$ being favored, which also implies very small opening angles in the diphoton system.

A recent study explored the capability of the gaseous argon MPD to search for decays of dark sector particles, including dark photons and dark Higgs bosons that can decay fully visibly into the final state $e^+ e^-$~\cite{Berryman:2019dme}. This background channel has a decent degree of overlap with the $a \to \gamma\gamma$ channel we are interested in because its dominant background is from the NC$\pi^0$ production. \footnote{In contrast to the search presented here, for NC$\pi^0$ events (with $\pi^0 \to \gamma\gamma$) to contribute to backgrounds like $A^\prime \to e^+ e^-$, one of the final-state photons must be misidentified or too low-energy to be detected. Here, we require that both photons are identified. Ref.~\cite{Berryman:2019dme} estimated that 10\% of photons in the NC$\pi^0$ sample are missed.} The searches in Ref.~\cite{Berryman:2019dme} involved lower-energy new-physics particles than those compared here, so the high energy of $E_{\gamma\gamma}$ provides an additional mechanism to separate our signal from the NC$\pi^0$ backgrounds. Therefore, we expect that a nearly background-free search for $a \to \gamma\gamma$ is possible and will proceed under that assumption.

If we shift our focus to the hadronic final states, $a \to$ hadrons, the signal characteristics are not too different. Especially if we consider the $a$ decay as the process $a \to$ hadrons, we can characterize the final state in terms of the total hadronic energy $E_{\rm had.}$ and an opening angle $\Delta\theta_{\rm had.}$, which is a proxy for the total jet size of all of the final state hadrons.
\begin{figure}
\begin{center}
\includegraphics[width=0.485\linewidth]{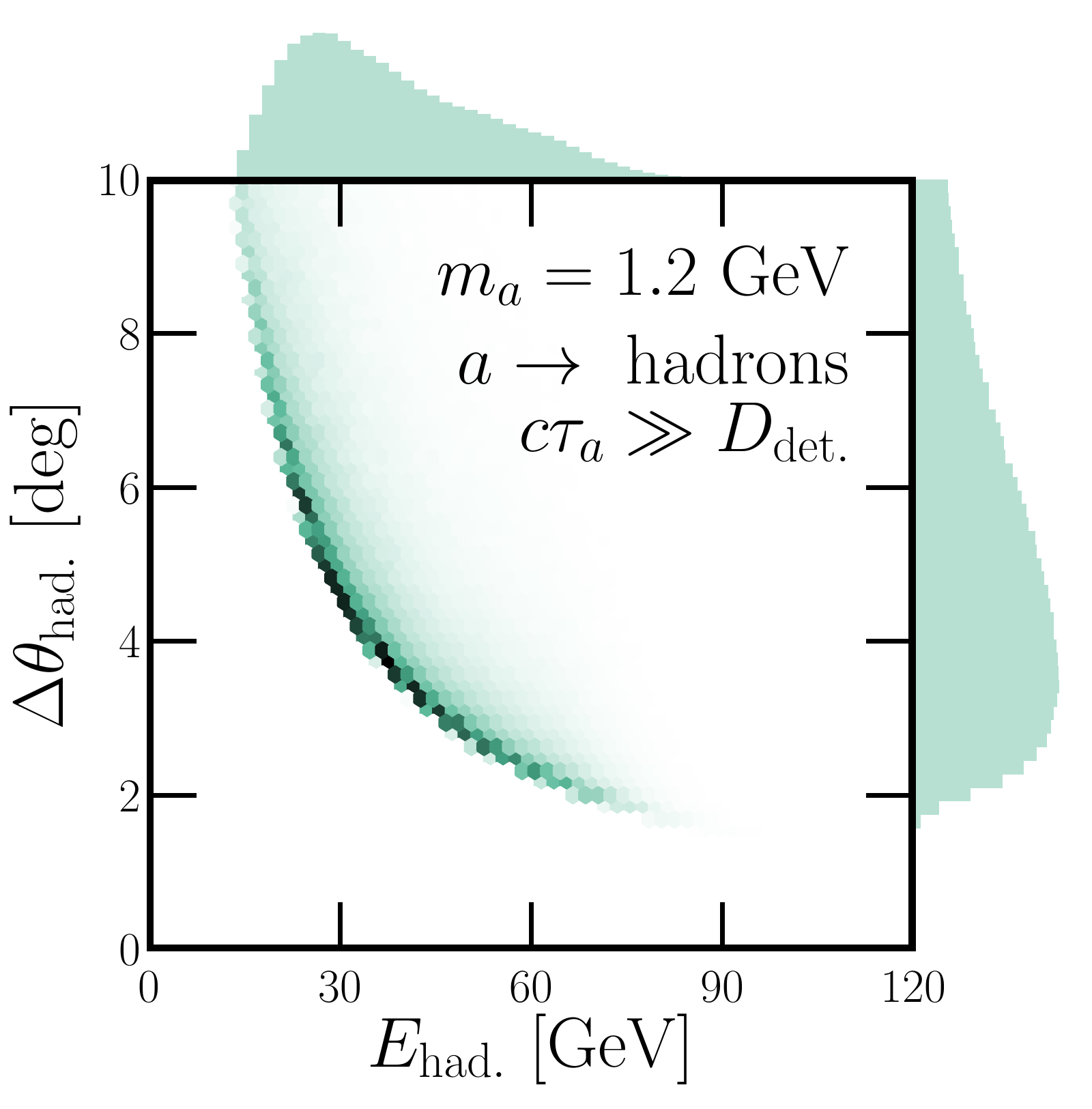}
\includegraphics[width=0.485\linewidth]{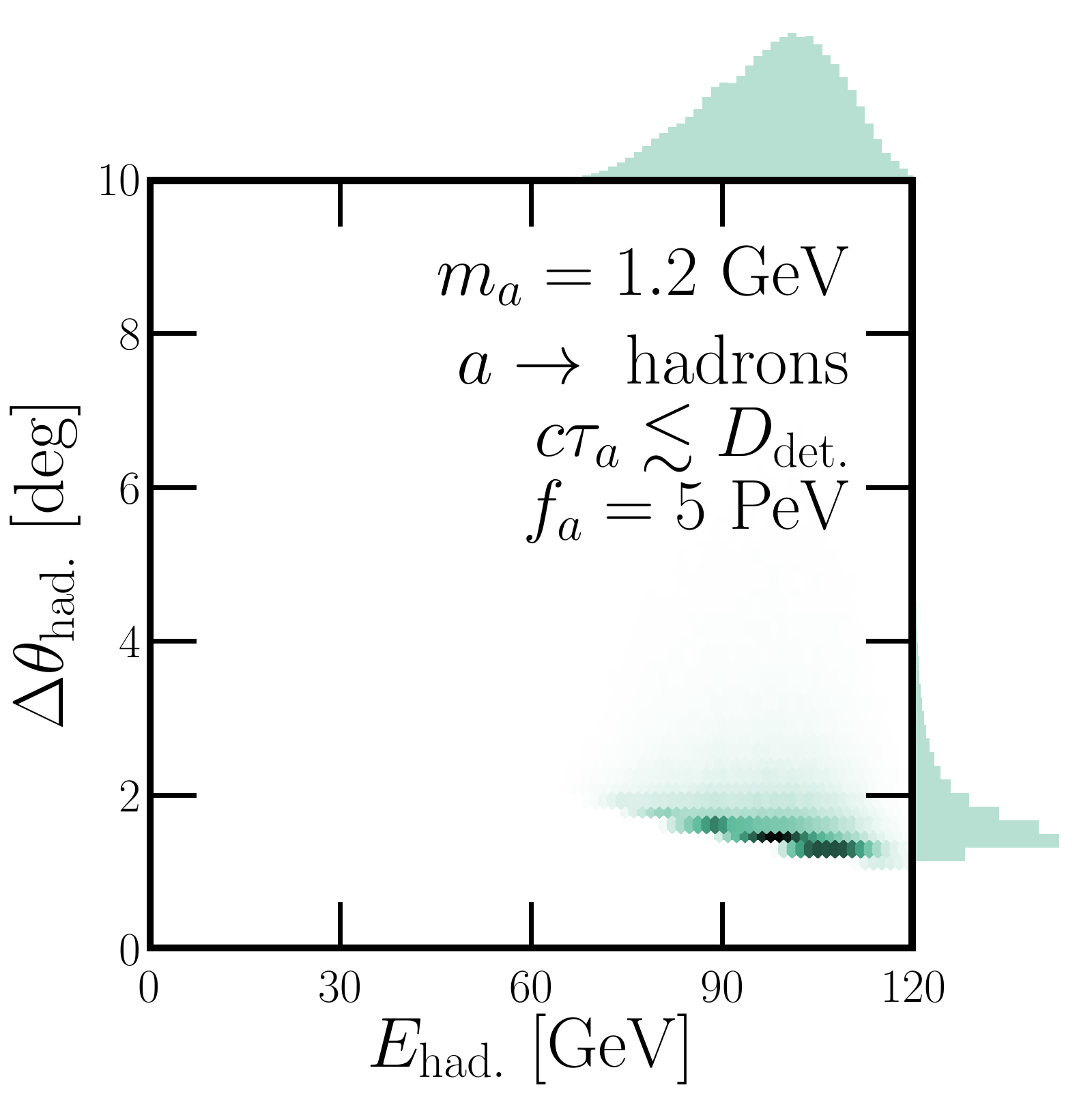}
\caption{Signal distributions of the decay $a\to$ hadrons, which we characterize using the observables $E_{\rm had.}$ and $\Delta\theta_{\rm had.}$, which gives the size of the hadronic jet in the final state. Here, $m_a = 1.2$ GeV and $a$ is produced via gluon-gluon fusion. Like in Fig.~\ref{fig:2DDist_MesMix}, the left (right) panel corresponds to a situation in which $a$ is long-(short-)lived relative to the distance between the DUNE target and ND. One-dimensional distributions for the two observables are shown on top of/to the right of the two-dimensional distribution panels.
\label{fig:2DDist_ggF}}
\end{center}
\end{figure}
Signal distributions of this variety are shown in Fig.~\ref{fig:2DDist_ggF}, where we now assume that $m_a = 1.2$ GeV and that $a$ is produced via the gluon-gluon fusion process discussed above. As with Fig.~\ref{fig:2DDist_MesMix}, we display the event distributions with respect to $E_{\rm had.}$ and $\Delta\theta_{\rm had.}$.\footnote{The opening angle is calculated assuming a two-body final state, a good description for the $a\to gg$ decay.} The left (right) panel assumes that $a$ is long-(short-)lived relative to the distance between the DUNE target and ND. This explains why lower energies are favored in the left panel and higher energies in the right one. We note that the one-dimensional $E_{\rm had.}$ distributions in Fig.~\ref{fig:2DDist_ggF} on top of each panel nearly match the shapes of the histograms in Fig.~\ref{fig:ggFdist} for the ``at Production'' and $f_G = 50$ PeV choices.

\begin{figure}
\begin{center}
\includegraphics[width=0.6\linewidth]{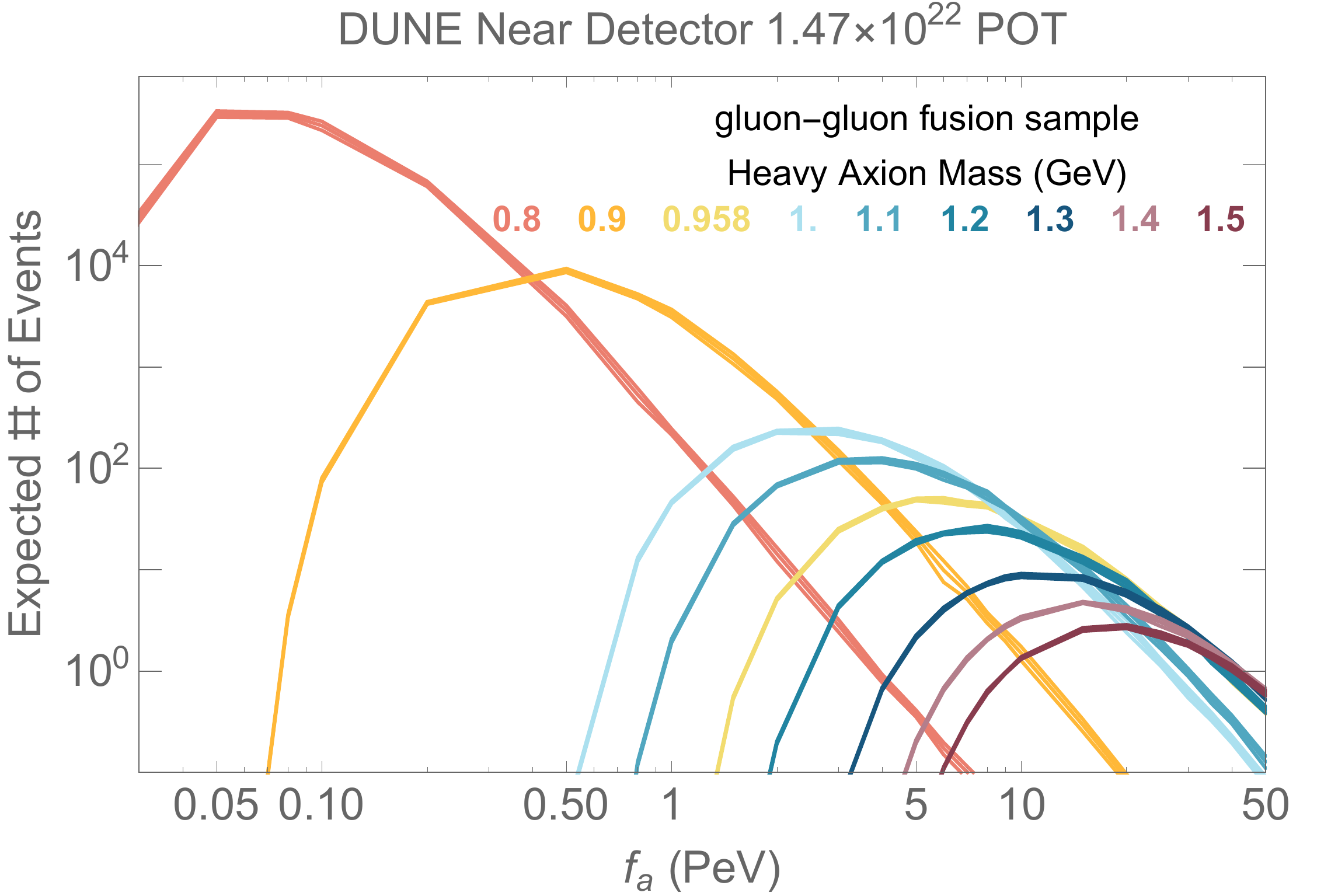}
\caption{The number of expected signal events for various heavy axion masses as a function of the axion decay constant $f_a$ from the gluon-gluon fusion process. The multiple lines with the same color for each $m_a$ represent the numerical uncertainty of our simulation.
\label{fig:ggFeff}}
\end{center}
\end{figure}

In Fig.~\ref{fig:ggFeff}, we show the number of expected signal events as a function of the axion decay constant $f_a$ for various axion mass points for the gluon-gluon fusion sample. For each mass point, we draw three curves of the same color to indicate the numerical uncertainties of our study. For low decay constant $f_a$, the production rate is high, but the detection probability is exponentially suppressed by the detector distance $D$ over the lab-frame decay probability, $\exp(-D/\beta\gamma c\tau)$. For high decay constant $f_a$, the lifetime is long, and the expected signal number is suppressed by $1/f_a^2$ for production rate and $L/(\beta\gamma c\tau)$ for the detection probability. Here $L$ is the effective detector length in the line-of-flight for the axion. Due to the large spread of the axion boost factors at production, the transition between these two limits spreads over the decay constant over a decade or so. We can see, as anticipated, the expected number of signal events decreases for increasing mass due to the rate suppression. More importantly, a larger signal mass means a smaller boost, a shorter lifetime. To reach the DUNE ND, it requires larger $f_a$ to overcome the arrival flux suppression $\exp(-D/\beta\gamma c\tau)$. Similar suppression exists for resonant mixing, as we show in yellow color when the axion mass is nearly degenerate with $\eta^\prime$ meson. Nevertheless, thanks to the large flux at DUNE, the DUNE ND will be able to probe the high $f_a$ regime uniquely, as shown in the next section.
%This implies that for heavy axion searches, nearer detectors might be able to catch more of the flux.
%For each mass, there is an optimal decay constant $f_a$ value, it roughly corresponds to the optimal

Backgrounds in this channel are mainly from scattering events that produce many final-state pions, etc., including charged-current scattering that produces a single charged muon and one or more pions ($\text{CC}1\mu2\pi$). Deeply-inelastic-scattering (DIS) events, where the argon nucleus is completely broken up, can also result in events that would mimic this signal. However, as all of these events are generated by neutrino scattering, their total energy will (as in the NC$\pi^0$ case) be less than roughly $E_{\rm had} \lesssim 5$ GeV. Our signal events, as demonstrated by Fig.~\ref{fig:2DDist_ggF}, will have hadronic energies $\gtrsim30$ GeV, and even higher if the axion is short-lived. As with the $a \to \gamma\gamma$ final state, the direction of these events is very forward-going, whereas the background will be more isotropic, and the opening angle is much smaller in the signal distributions than the backgrounds. With all of these features, we expect the $a\to$ hadrons search channel to be background free, like the $a\to\gamma\gamma$ channel.

Before proceeding, we also wish to discuss one unique strength of the search at DUNE ND: combining searches for decaying heavy axions in both the liquid and gas detectors into one combined analysis. The background contributions discussed above are from beam neutrinos scattering in one of the detectors. These background rates scale with detector mass, and so the expected background contributions in the liquid detector are a factor of over $50$ higher than in the gaseous detector. Meanwhile, the signal rate of decaying axions is more-or-less proportional to the volume of the detector and, therefore, will be roughly equal in the two detectors. A combined analysis, where the expected signal-to-background ratio can be robustly predicted from one detector to the other, can improve the overall DUNE ND capability.
\section{DUNE Near Detector Sensitivity to Heavy Axions}\label{sec.NDsens}
Combining all of the ingredients discussed to this point, we are now prepared to estimate the DUNE ND sensitivity to heavy axions.
%We first show the results and discuss in details the results in the in gluon dominance scenario, and then show the results for the codominance scenario, where the scenarios are defined in Sec.~\ref{subsec:EFT}.

\subsection{Gluon Dominance Projections}\label{subsec:GD}
\begin{figure}
\begin{center}
\includegraphics[width=0.9\linewidth]{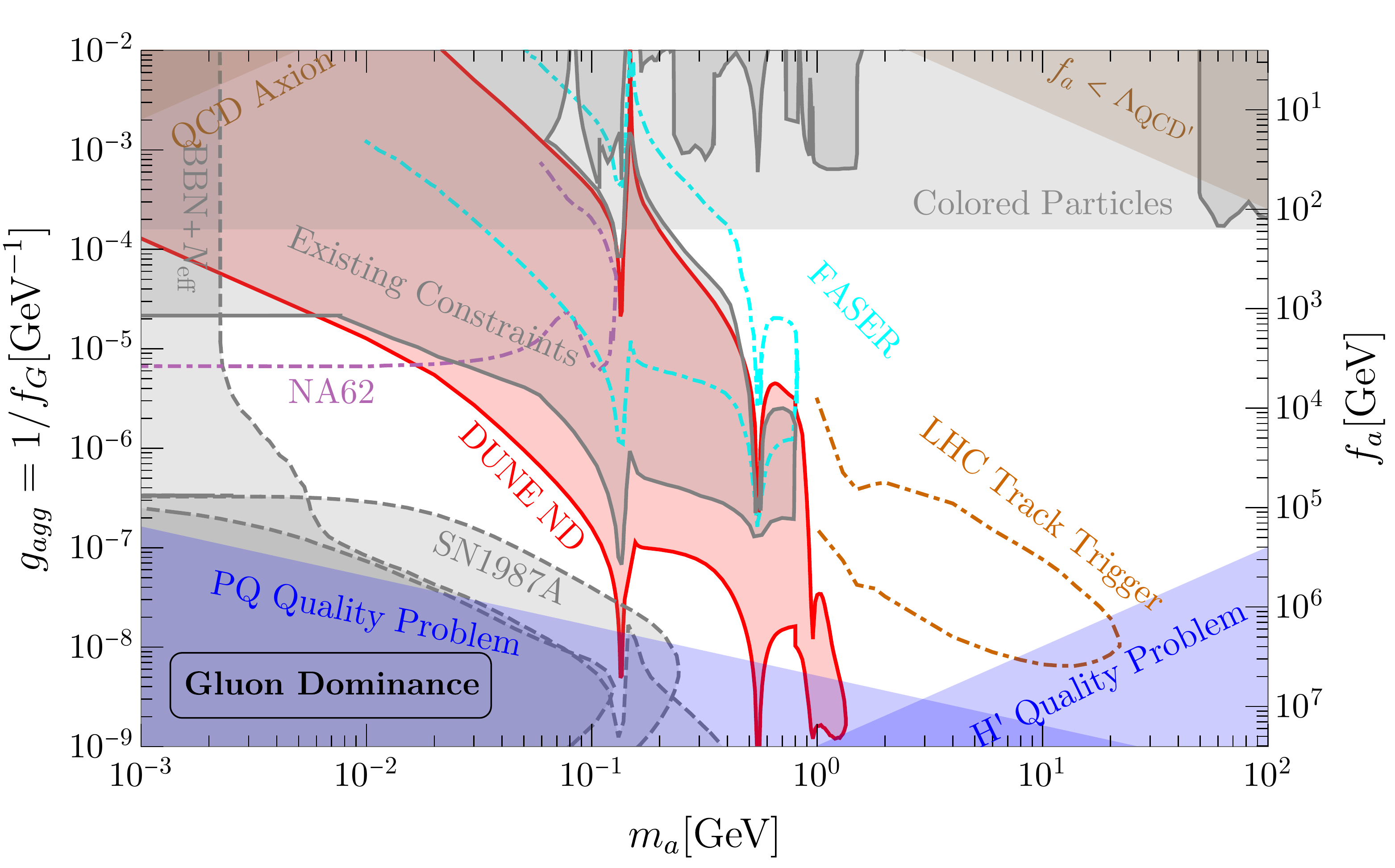}
\caption{Expected sensitivity at DUNE Near Detector (in red) for ``Gluon Dominance'' scenario with $c_3=1, c_1, c_2=0$ in Eq.~\eqref{eq:alpeft} along with existing constraints and coverage by future experiments. The constraints from SN1987A \cite{Chang:2018rso,Ertas:2020xcc} and cosmology \cite{Depta:2020wmr} are shown in dashed lines given the astrophysical uncertainties and model dependence. The region ``Existing Constraints'' include the bounds from partially invisible kaon decays from E787 and E949~\cite{Ertas:2020xcc}, electron beam dump~\cite{Dolan:2017osp,Dobrich:2015jyk,Banerjee:2020fue}, CHARM~\cite{Ariga:2018uku}, visible kaon decays~\cite{Gori:2020xvq}, B decays \cite{Aloni:2018vki}, LHC dijet searches~\cite{Mariotti:2017vtv}. We also include projections relying on the proposed displaced track trigger at the HL-LHC~\cite{Hook:2019qoh}, FASER~\cite{Ariga:2018uku} and NA62~\cite{Ertas:2020xcc}.
\label{fig:DUNESensitivity}}
\end{center}
\end{figure}
We first focus on the case of gluon dominance discussed in Sec.~\ref{subsec:EFT} where $c_3=1, c_1, c_2=0$.\footnote{There would still be a non-negligible, $c_3$-induced photon coupling even with $c_1=c_2=0$.} 
Combining both the meson mixing and gluon-gluon fusion production modes, we determine the parameter space for which we would expect three or more signal events in ten years of data collection at DUNE ND, the red shaded region in Fig.~\ref{fig:DUNESensitivity}. The blue shaded regions correspond to the same ones shown in Fig.~\ref{fig:theoryspace} based on theoretical considerations regarding the axion Quality Problem. The brown shaded regions are the same as in Fig.~\ref{fig:theoryspace} as well. The horizontal shaded region labeled ``Colored Particles'' is disfavored since from a UV perspective, $aG\tilde{G}$ coupling generically originate after integrating out colored fermions with masses $\sim y f_a$. Requiring a maximal Yukawa coupling $y\sim 4\pi$ along with the LHC constraints on colored states to have masses above $2$~TeV~\cite{Aaboud:2017nmi,Sirunyan:2018xlo,Sirunyan:2018rlj,Aad:2019hjw} gives the above bound.

We also include a number of existing experimental/observational constraints on this parameter space in grey. \footnote{Here we update some of the astrophysical and cosmological bounds used in Ref.~\cite{Hook:2019qoh} by using the more recent results from~\cite{Depta:2020wmr,Ertas:2020xcc}.} For small mass $m_a\lesssim100$~MeV, astrophysical and cosmological constraints are relevant --- the region labeled ``SN1987A'' indicates the region of parameter space for which such axions would cool the supernova and carry away too much energy~\cite{Chang:2018rso,Ertas:2020xcc},\footnote{To illustrate the uncertainty of the SN1987A bound, we show two contours corresponding to fiducial profiles used in each of~\cite{Chang:2018rso} and \cite{Ertas:2020xcc}.}\footnote{For a recent update on SN1987A bound on ALP-photon coupling, see~\cite{Lucente:2020whw}.} whereas the region labeled ``BBN + $N_{\rm eff}$'' indicates where such axions would affect light-element abundances and contribute to the number of effective degrees of radiation~\cite{Depta:2020wmr,Millea:2015qra}.\footnote{For a recent discussion on BBN constraint on ALP-lepton couplings see~\cite{Ghosh:2020vti}.} Both the ``SN1987A'' bound and ``BBN + $N_{\rm eff}$'' are shown via dashed lines because of their associated uncertainties, see e.g.~\cite{Bar:2019ifz} and~\cite{Depta:2020wmr}, respectively. For $m_a$ between 1 MeV and 1 GeV, a number of searches have been performed in the contexts of both electron %(E137~\cite{Bjorken:1988as}, E141~\cite{Riordan:1987aw})
and proton beam dumps
%(CHARM~\cite{Bergsma:1985qz}, NuCal~\cite{Blumlein:1990ay,Blumlein:1991xh})
and corresponding bounds were discussed in~\cite{Dobrich:2015jyk,Dolan:2017osp,Ariga:2018uku} for these types of axions, as well as searches for rare meson decays~\cite{Izaguirre:2016dfi,Aloni:2018vki,Gavela:2019wzg,Ertas:2020xcc,Gori:2020xvq}. Furthermore, there are constraints from LHC dijet searches for $m_a > 50$~GeV as obtained in~\cite{Mariotti:2017vtv}.

Other planned experiments with similar timescales as DUNE are capable of performing searches in this region of parameter space. We include some projections of these in Fig.~\ref{fig:DUNESensitivity} as well.\footnote{For comparison against other potential future experimental searches in this parameter space, see Fig.~\ref{fig:DUNESensitivityprop} in Appendix~\ref{app:Comparison}.} The FASER~\cite{Ariga:2018uku} (dot-dashed cyan) experiment at the LHC will be able to probe a similar mass regime as DUNE but with smaller $f_G$ due to its close proximity/high-energy production source. A proposed displaced decay search using the high-luminosity LHC track trigger~\cite{Hook:2019qoh} can probe the region encompassed by the dot-dashed brown line at heavier masses than DUNE. Finally, for $m_a < m_{\pi}$ , NA62 has powerful sensitivity (dot-dashed purple) via the search for $K^+ \rightarrow \pi^++a$ where $a$ is undetected~\cite{Ertas:2020xcc}. Comparing our DUNE projections against these other future proposals, the complementarity of these different search strategies is obvious – the combination of all of these will allow for considerable reach in the theoretically- motivated parameter space in a way that no individual experiment can accomplish on its own. DUNE will specifically be most powerful in this long-lived region of small $g_{agg}$ or large $f_G$, especially for $20~\text{MeV}\lesssim m_a\lesssim 2~\text{GeV}$.

\subsection{Codominance Projections}
\begin{figure}
\begin{center}
\includegraphics[width=0.9\linewidth]{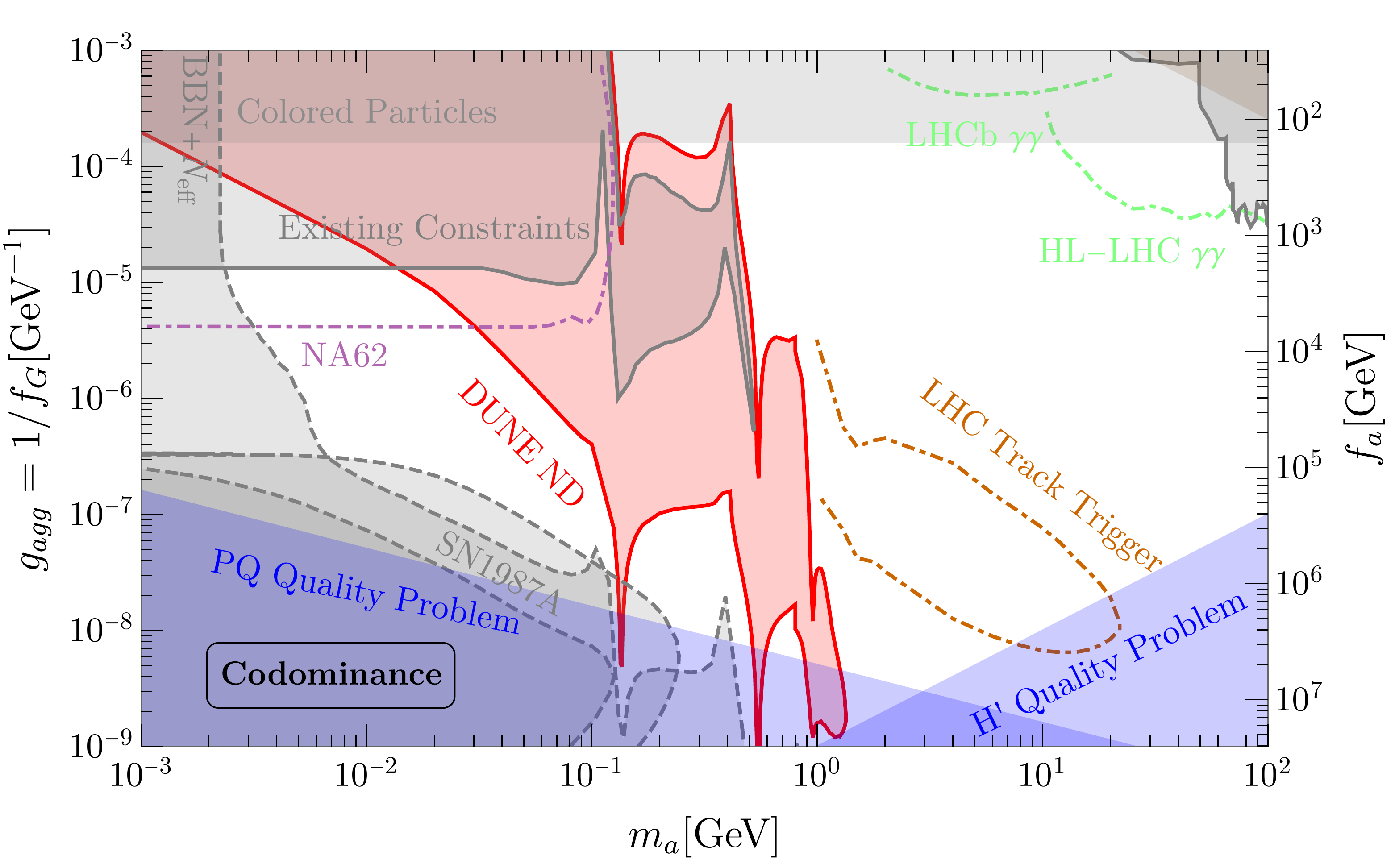}
\caption{Expected sensitivity at DUNE Near Detector (in red) for the ``Codominance'' scenario with $c_1=c_2=c_3=1$ in Eq.~\eqref{eq:alpeft} along with existing constraints and coverage by future experiments. The constraints from SN1987A~\cite{Chang:2018rso,Ertas:2020xcc} and cosmology~\cite{Depta:2020wmr} (appropriately recasted for our purpose) are shown in dashed lines given the astrophysical uncertainties and model dependence. The region ``Existing Constraints'' include the bounds from kaon decays from E787 and E949~\cite{Ertas:2020xcc,Gavela:2019wzg}, electron beam dump~\cite{Dolan:2017osp,Dobrich:2015jyk,Banerjee:2020fue}, as well as diphoton and dijet searches at the LHC~\cite{Mariotti:2017vtv}. We also include projections relying on the proposed displaced track trigger at the HL-LHC \cite{Hook:2019qoh}, NA62~\cite{Ertas:2020xcc} and diphoton searches at LHCb~\cite{CidVidal:2018blh} and HL-LHC~\cite{Mariotti:2017vtv}. For the present ``Codominance'' scenario, we have not reanalyzed the coverage discussed in~\cite{Gori:2020xvq} for NA62, NA48, KOTO and in~\cite{Gavela:2019wzg} for Belle-II. They would cover a mostly complementary region roughly for $1/f_G>10^{-4}\text{GeV}^{-1}$ and $m_a\lesssim \text{few GeV}$. The existing CHARM data~\cite{Bergsma:1985qz} would also cover some part of the parameter space for roughly $100~\text{MeV}< m_a < 1~\text{GeV}$ (details in the text).\label{fig:DUNESensitivityCo}}
\end{center}
\end{figure}
Here we focus on the scenario where $c_1 = c_2 = c_3$ and derive the coverage for 3 events for 10 year data taking at DUNE ND, shown in red in Fig.~\ref{fig:DUNESensitivityCo}. While this coverage is similar to the one in Fig.~\ref{fig:DUNESensitivity}, around $m_a \sim 400$~ MeV, the effective photon coupling $c_\gamma$, in Eq.~\eqref{eq.cgalow} becomes small for our choices of $c_i$. As a result, the coverage region shifts upwards exhibiting a “peak”-like feature.

The theoretical constraints from the axion Quality Problem remain the same. The set of experimental/observational constraints are shown in grey. Since the ``SN1987A'' constraints are dominated by the $G\tilde{G}$ coupling, it remains the same. However, the ``BBN+$N_{\rm eff}$'' constraint is dominated by the $F\tilde{F}$ coupling, and hence it gets modified based on Eq.~\eqref{eq.cgalow} as a result of non-negligible $c_1$ and $c_2$. The constraints from partially invisible kaon decays are also modified due to the non-negligible $aW\tilde{W}$ coupling \cite{Izaguirre:2016dfi,Gavela:2019wzg,Ertas:2020xcc}. We recast the electron and proton beam dump results from~\cite{Dolan:2017osp} as appropriate for the present case of Codominance. Since the diphoton decay modes are now non-negligible, for higher masses $m_a > 20$ GeV, both the diphoton and dijet searches at the LHC give relevant constraints~\cite{Mariotti:2017vtv}.

We also include future projections from diphoton searches at LHCb~\cite{CidVidal:2018blh} and HL-LHC~\cite{Mariotti:2017vtv} (dot-dashed green), kaon decay searches at NA62~\cite{Gavela:2019wzg,Ertas:2020xcc} (dot-dashed purple) and LHC Track Trigger proposal~\cite{Hook:2019qoh} (dot-dashed brown) can cover complementary regions of parameter space as before.

Some comments regarding a few omissions in Fig.~\ref{fig:DUNESensitivityCo} are in order. We expect some part of the parameter space for $100~\text{MeV}< m_a < 1~\text{GeV}$ would be covered by the existing CHARM data~\cite{Bergsma:1985qz} which we have not derived for the Codominance scenario. Also, we have not derived the constraints and projections from KOTO and NA62/48 from visible kaon decays for this scenario, which would cover some parameter space for $150~\text{MeV}< m_a < 350~\text{MeV}$ and roughly $1/f_G>10^{-4}~\text{GeV}^{-1}$, mostly complementary to our DUNE ND coverage. In Ref.~\cite{Gori:2020xvq}, such constraints were derived for the cases $G\tilde{G}$ and $W\tilde{W}$ -dominance separately. Some complementary coverage, again for roughly $1/f_G>10^{-4}~\text{GeV}^{-1}$ and $m_a\lesssim\text{few GeV}$, would also come from $B\rightarrow Ka$ processes at Belle-II similar to what is discussed in~\cite{Gavela:2019wzg} for the case of $W\tilde{W}$-dominance.

To summarize, similar to the case of ``Gluon Dominance'' above, we see that DUNE ND would provide a powerful coverage, complementary to other existing and projected constraints, especially for large $f_G$ and $30~\text{MeV}\lesssim m_a\lesssim 1~\text{GeV}$.

\section{Conclusion}\label{sec.conclusion}
Recent studies of the Strong CP Problem (and the associated Axion Quality Problem) have led to a renewed interest in heavy axions with masses in the MeV-TeV regime. Meanwhile, a number of upcoming and planned experiments are capable of searching for decays of long-lived particles in a beam dump environment. One of the best example, in terms of the total protons on target (POT) and large, multipurpose detectors, of such an experiment is the Deep Underground Neutrino Experiment (DUNE) with its Near Detector (ND) complex. Combining the intense, high-energy proton beam (with a large number of POT per year) and the fine-grained NDs (both the liquid and the gaseous argon ones, allowing for particle identification and energy resolution) provides an exciting prospect for such searches.

In this paper, we have thoroughly explored the DUNE ND complex's ability to search for heavy axions in the MeV-GeV regime. We have revisited previous considerations of heavy axion production through both neutral, pseudoscalar mixing as well as through gluon-gluon fusion. Motivated by the Strong CP Problem, we have focused on two cases of these heavy axions via an Effective Field Theory treatment -- one where the axion's dominant coupling is to the SM gluon field strength tensor, and one where it couples democratically to each of the SM gauge group field strength tensors. This is a different focus than the often-studied photon-dominant scenario for axion-like particle searches in beam dump environments. %, where such production processes through meson mixing and gluon-gluon fusion are less relevant.

The DUNE NDs offer several ways of identifying the decays of these heavy axions in their dominant decay channels, which are, depending on the axion mass, into photon pairs or hadrons. We have identified how these searches can leverage different signal characteristics to fully suppress neutrino-related backgrounds, allowing for very powerful searches of these rare signatures. Comparing to other projections for these classes of heavy axions, DUNE provides complementary sensitivity, specifically to very long-lived axions. Performing this type of search in tandem with other collider-based or meson-decay-based searches will allow us to cover as much of the theoretically-motivated parameter space as possible. There exist many more ways to explore these intriguing heavy axion theories at DUNE, including a large variety of production modes, from bremsstrahlung, meson decays mediated by operators beyond the gluon field strength, meson flavor changing decays, hadronic Primakov processes, as well as the rich decay channels from different Axion EFTs.

Whether or not an axion exists as a solution to the Strong CP Problem, as well as if it is in this heavy-axion category, remains to be seen. Regardless, experiments such as DUNE can perform unique searches for these and other new-physics scenarios without detracting from their overall scientific missions (in this case, neutrino oscillation studies). It is imperative that these searches are performed so that our planned experiments can extract as much scientific knowledge as they can. If such a heavy axion does exist within the reach of DUNE ND, then not only will DUNE revolutionize the field of neutrino physics, it will revolutionize our understanding of axions as well.

\acknowledgements{We thank Hsin-Chia Cheng,  Fatih Ertas,  Jan Jerhot,  Felix Kahlhoefer, Gustavo Marques-Tavares and Dean Robinson for helpful discussion. KJK is supported by Fermi Research Alliance, LLC, under contract DE-AC02-07CH11359 with the U.S. Department of Energy. SK was supported in part by the NSF grants PHY-1914731, PHY-1915314 and the U.S. DOE Contract DE-AC02-05CH11231. ZL was supported in part by the NSF grants PHY-1620074, PHY-1914480, and PHY-1914731, and by the Maryland Center for Fundamental Physics (MCFP). The code involved in this study are available at ~\href{https://gitlab.com/ZhenLiuPhys/alpdune}{\tt GitLab}.}

%\clearpage
\appendix
\section{Comparison Against Other Proposed Experiments}\label{app:Comparison}

To place our projected limits on the gluon dominance scenario discussed in Section~\ref{subsec:GD}, here we provide a version of Fig.~\ref{fig:DUNESensitivity} with a larger set of future, proposed experimental sensitivities. In addition to the projections from NA62, FASER, and the HL-LHC shown and discussed in the main text, we include here projections from FASER2~\cite{Feng:2018noy,Ariga:2018uku}, CODEX-b~\cite{Gligorov:2017nwh}, and MATHUSLA~\cite{Chou:2016lxi}. Given the energies/detector locations of these different proposals, we see that, DUNE will still have unique sensitivity at large $f_G$/small $g_{agg}$ as long as $20$ MeV $\lesssim m_a \lesssim 2$ GeV. These other proposals, specifically CODEX-b and MATHUSLA, provide sensitivity at higher $m_a$ in the same region of parameter space as the LHC Track Trigger proposal~\cite{Hook:2019qoh}, $1$ GeV $\lesssim m_a \lesssim 10$ GeV, another interesting regime for heavy axion searches. We note here that the MATHUSLA and CODEX-b projections are taken from a recent analysis in Ref.~\cite{Aielli:2019ivi}, where new production modes from gluon splitting, gluon-gluon fusion, and meson decays are included. 
\begin{figure}
\begin{center}
\includegraphics[width=0.9\linewidth]{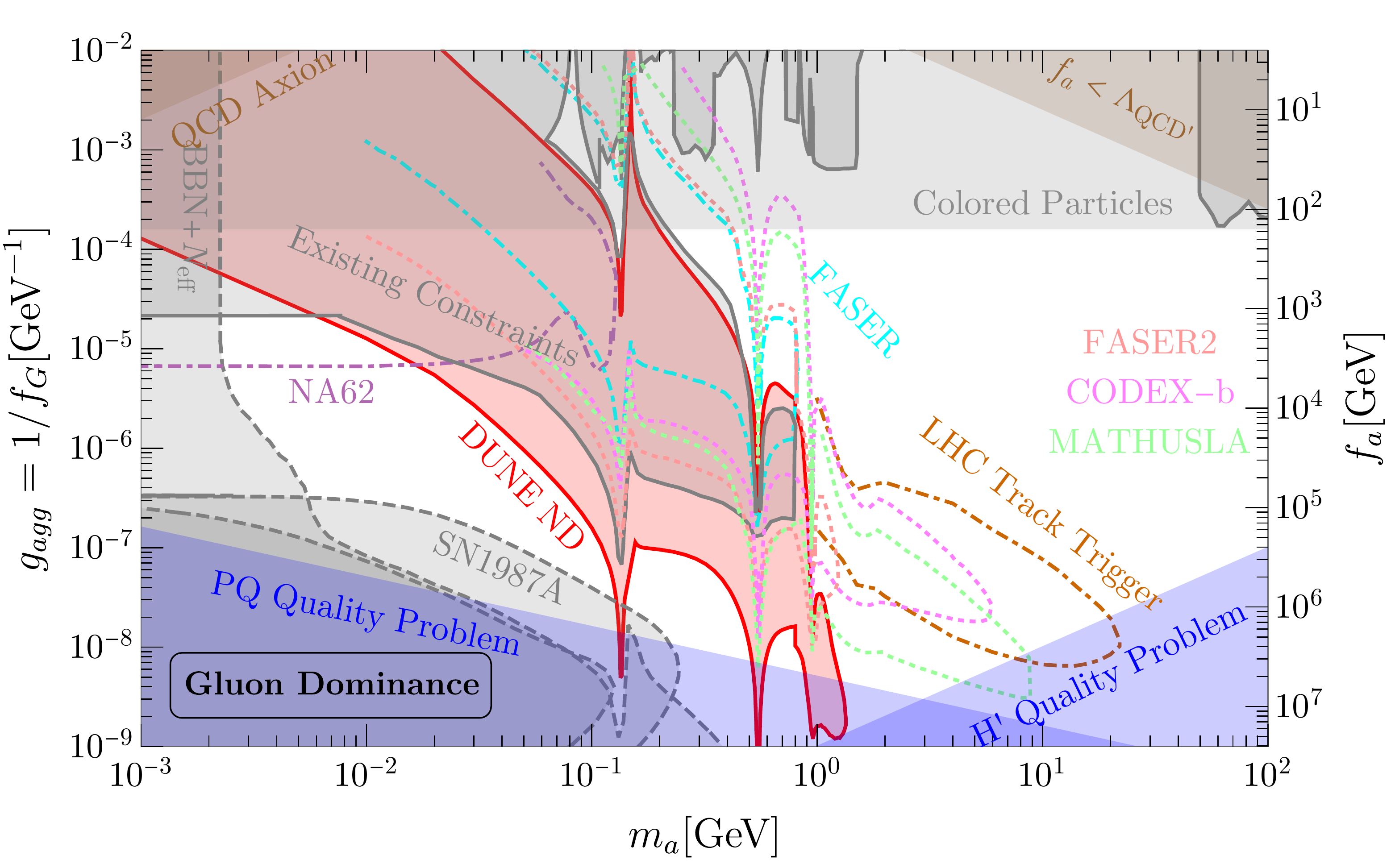}
\caption{%Expected sensitivity at DUNE Near Detector (in red) for the ``Gluon Dominance'' scenario along with existing constraints and coverage by future and proposed experiments.
Same as Fig.~\ref{fig:DUNESensitivity} along with complementary, projected coverage by FASER2~\cite{Feng:2018noy,Ariga:2018uku}, MATHUSLA~\cite{Chou:2016lxi} and CODEX-b~\cite{Gligorov:2017nwh} for the Gluon Dominance scenario. The latter two projections on MATHUSLA and CODEX-b are taken from a recent analysis in Ref.~\cite{Aielli:2019ivi}.
% where the coverage are subject to correction (see details in the text).
\label{fig:DUNESensitivityprop}}
\end{center}
\end{figure}

\bibliographystyle{utphys}%{utphys}%{plainnat}
\bibliography{refs_Axion}

\end{document}